\documentclass[11pt]{article}
\usepackage{comment}
\usepackage[margin=2.5cm]{geometry}
\usepackage{amsmath,amssymb,extarrows,mathtools,graphicx,subfigure,setspace,bbold}
\usepackage{cite}
\usepackage{slashed}
\usepackage[dvipsnames]{xcolor}
\usepackage{hyperref}
\hypersetup{colorlinks=true, linkcolor=blue, citecolor=magenta, linktoc=page}
\usepackage{tikz}
\usepackage{arydshln,leftidx,mathtools}
\usetikzlibrary{decorations.pathreplacing}
\numberwithin{equation}{section}

\DeclareMathOperator{\Tr}{Tr}

\def\({\left(}
\def\){\right)}

\newcommand{\be}{\begin{equation}}
\newcommand{\ba}{\begin{eqnarray}}
\newcommand{\ea}{\end{eqnarray}}
\newcommand{\ee}{\end{equation}}

\newcommand{\bal}{\begin{align}}
\newcommand{\eal}{\end{align}}
\newcommand{\bes}{\begin{equation*}}
\newcommand{\bas}{\begin{eqnarray*}}
\newcommand{\eas}{\end{eqnarray*}}
\newcommand{\ees}{\end{equation*}}
\newcommand{\bals}{\begin{align*}}
\newcommand{\eals}{\end{align*}}

\begin{document}

\begin{titlepage}
\thispagestyle{empty}

\begin{flushright}
IPM/P-2015/051\\
\end{flushright}

\vspace{.4cm}
\begin{center}
\noindent{\Large \textbf{Holographic Mutual Information for Singular Surfaces}}\\
\vspace{2cm}

M. Reza Mohammadi Mozaffar$\, ^{\dag}$, Ali Mollabashi$\, ^{\dag}$ and Farzad Omidi$\, ^{\ddag}$
\vspace{1cm}

$^\dag$ {\it School of Physics,} $^\ddag$ {\it School of Astronomy,}
\\
{\it Institute for Research in Fundamental Sciences (IPM), Tehran, Iran}\\
\vspace{1cm}
Emails: {\tt m$_{-}$mohammadi, mollabashi, and farzad@ipm.ir}

\vskip 2em
\end{center}

\vspace{.5cm}
\begin{abstract}
We study corner contributions to holographic mutual information for entangling regions composed of a set of disjoint
sectors of a single infinite circle in 3-dimensional conformal field theories. In spite of the UV divergence of holographic mutual information, it exhibits a first order phase transition.
We show that tripartite information is also divergent for disjoint sectors, which is in contrast with the well-known feature of tripartite information being finite even when entangling regions share boundaries.
We also verify the locality of corner effects by studying mutual information between regions separated by a sharp annular region. 
Possible extensions to higher dimensions and hyperscaling violating geometries is also considered for disjoint sectors. 
%We also discuss about constructing a universal quantity from mutual information of contacting sharp entangling regions.
\end{abstract}

\end{titlepage}

\newpage

%\begin{scriptsize}
\tableofcontents
\noindent
\hrulefill
%\end{scriptsize}

%\newpage

\onehalfspacing

%\begin{tikzpicture}[scale=6]
%  \filldraw[fill=green!20!white, draw=green!50!black]
%    (90:3mm) arc (90:0:3mm) arc (180:90:3mm) ;
%\end{tikzpicture} 

%\begin{tikzpicture}[scale=6]
%  \filldraw[fill=green!20!white, draw=green!50!black]
%    (90:3mm) arc (90:0:3mm) arc (180:90:3mm) -- (0:6mm) -- (0,0) -- cycle;
%\end{tikzpicture}  
\section{Introduction}
Quantum entanglement is one of the important features of quantum mechanics that emerges in several areas of physics including condensed matter, quantum information and black hole physics.
In the context of quantum field theories (QFTs), the notion of entanglement can reflect various features of the theory, depending on how the Hilbert space of the theory is decomposed. This leads to different types of entanglement in QFTs including those which are denoted in the literature by spatial (or geometric) entanglement \cite{Bombelli:1986rw, Callan:1994py, Srednicki:1993im, Calabrese:2004eu}, momentum space entanglement \cite{Balasubramanian:2011wt}, and field space entanglement \cite{Yamazaki:2013xva, Taylor:2015kda, Mozaffar:2015bda}.
%\footnote{In the context of quantum field theory we can consider various kind of entanglement between different degrees of freedom which depends on the partitioning of the Hilbert space, e.g., field space entanglement, momentum space entanglement and spatial (or geometric) entanglement \cite{{Yamazaki:2013xva},{Taylor:2015kda},{Mozaffar:2015bda},{Balasubramanian:2011wt},{Callan:1994py}}.
Here we focus on geometric entanglement, which has been widely studied during recent years, in order to extend some recent developments in the context of corner contributions to different entanglement measures.

Entanglement entropy is a well-known measure to quantify quantum entanglement. In the context of geometrical entanglement, in order to define entanglement entropy (EE) we consider a spatial region $V$ on a constant time slice of a $d$ dimensional field theory. Assuming that the total Hilbert space has a decomposition as $\mathcal{H}_{\rm{tot.}}=\mathcal{H}_{V}\otimes \mathcal{H}_{\bar{V}}$, one can define a reduced density matrix corresponding to region $V$ by integrating out the degrees of freedom living in its geometrical complement $\bar{V}$, i.e., $\rho_V=\Tr_{\bar{V}}\rho_{\rm{tot.}}$. Entanglement entropy is then defined as the von Nuemann entropy for this reduced density matrix which is given by $S_{EE}=-\Tr \rho_V \log \rho_V$.

Entanglement entropy in QFTs is a UV-divergent quantity which its leading divergence is proportional to the area of the entangling region $V$, due to the leading contribution of the near-boundary local degrees of freedom i.e. \cite{Bombelli:1986rw, Srednicki:1993im, Casini:2009sr}
\begin{align}\label{area}
S_{EE}= c^{(d-2)}\frac{\mathcal{A}_V}{\epsilon^{d-2}}+\cdots+s_{\rm univ.}^{(d)}+\mathcal{O}\left(\frac{\epsilon}{\ell}\right),\;\;\;\;\;s_{\rm univ.}^{(d)}=\begin{cases}
c_e^{(d)} \log \frac{\ell}{\epsilon}&~~ d:\rm{even}\\
c_o^{(d)} &~~ d:\rm{odd}
\end{cases}.
\end{align}
In the above expression $1/\epsilon$ is the UV cut-off, $\mathcal{A}_V$ is the area of the entangling surface, $\ell$ is the characteristic length of $V$, and $s_{\rm univ.}^{(d)}$ encodes some universal information about the QFT. Note that by varying $\epsilon$ the two constants $c_e^{(d)}$ and $c_o^{(d)}$ does not change and they are called universal in this sense. The well-known example of this universal information occurs in two dimensional conformal field theories where $c_e^{(2)}$ is proportional to the central charge of the theory \cite{Calabrese:2004eu}. 

Although the computation of EE even in the simplest case of two dimensional conformal field theories is not an easy task, for the case of field theories supporting a gravity dual in the context of AdS/CFT correspondence \cite{Maldacena:1997re}, one can apply the RT prescription \cite{Ryu:2006bv} to calculate the holographic entanglement entropy (HEE). This set up is based on finding a minimal area surface in the bulk of an asymptotically AdS geometry. This surface anchors on the boundary of $V$ on the conformal boundary of the bulk and the corresponding HEE is given by
\begin{align}
S_{EE}=\frac{\rm{min(Area)}}{4G_N}.
\end{align}
Using this prescription Eq.\eqref{area} is reproduced for strongly coupled conformal field theories supporting a gravitational dual \cite{Ryu:2006ef}.

\begin{figure}\label{fig:kc}
\begin{center}
\begin{tikzpicture}[scale=1.7]
\draw [fill=blue!20!white] (0,0)--(3,0)--(4,1)--(1,1)--(0,0);
\draw [fill=white] (1.25,0)--(2,.7)--(2.05,0)--(1.25,0);
\draw [blue!60!black,<->,line width=.2mm](1.74,.45).. controls (1.85,.4) and (1.93,.4) .. (2.02,.45);
\draw [blue!40!black] (1.8,0.2) node {{\large $\mathbf{\Omega}$}};
\draw [fill=blue!10!white] (6,0)--(7.7,.5)--(7.7,2.7)--(6,2.5)--(6,0);
\draw [fill=blue!10!white] (6,0)--(7.5,-.5)--(7.5,2.5)--(6,2.5)--(6,0);
\draw [blue!80!white,densely dashed,-] (6,0)--(7.5,.44);
\draw [fill=blue!20!white] (6,0)--(7.5,.44)--(7.5,2.5)--(6,2.5)--(6,0);
\draw [blue!60!black,<->,line width=.3mm](7.5,-.5).. controls (7.8,-.2) and (7.85,.4) .. (7.7,.5);
\draw [blue!60!black,thick] (7.95,0) node {{\large $\mathbf{\Omega}$}};
\end{tikzpicture}
\caption{Left: The blue plane represents a constant time slice of a $d=3$ CFT with a kink ($k$) entangling region on it. Right: A crease ($k\times R^m$) entangling region as a direct generalization of the kink in higher dimensions.}
\end{center}
\end{figure}
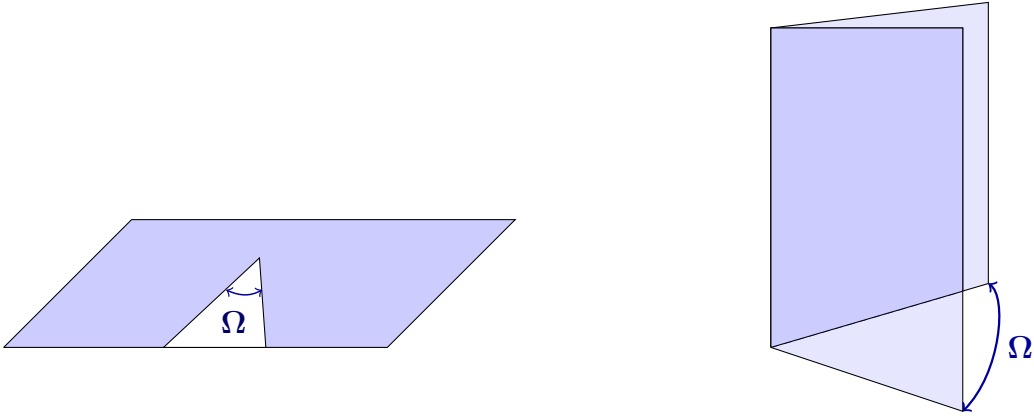

The structure of the universal terms appearing in the EE depends on the geometry of the entangling region. If the entangling region contains a singularity, it has been shown that extra divergent terms appear in the entanglement entropy \cite{Casini:2006hu, Casini:2008as, Hirata:2006jx, Myers:2012vs}. Two simple types of singular entangling regions which are known as kink (in $d=3$) and  crease (for $d\ge 4$) are shown in Fig.\ref{fig:kc}. The simplest example which is a kink in three dimensions with opening angle $\Omega$ and length $H$ has an extra logarithmic divergent term that modifies Eq.\ref{area} as
\begin{align}
S_{EE}=c^{(1)} \frac{\mathcal{A}_V}{\epsilon}-a(\Omega)\log \frac{H}{\epsilon}+c_0+\mathcal{O}\left(\frac{\epsilon}{H}\right),
\end{align}
where the function $a(\Omega)$ encodes some universal information about the theory and for a pure state it has a symmetric property as $a(\Omega)=a(2\pi-\Omega)$. The strong subadditivity property and Lorentz invariance of EE impose more constraints on $a(\Omega)$. In the large angle (smooth) limit and small angle (sharp) limit of the opening angle $\Omega$ one finds
\begin{align}
a(\Omega\rightarrow 0)=\frac{\kappa}{\Omega}+\cdots\;\;\;\;,\;\;\;\;a(\Omega\rightarrow \pi)=\sigma(\pi-\Omega)^2+\cdots,
\end{align}
where $\kappa$ and $\sigma$ correspond to some characteristics of the underlying CFT. In particular recently it was shown that the constant $\sigma$ is proportional to the central charge appearing in the two point function of the energy-momentum tensor as $\sigma=\frac{\pi^2}{24}C_T$, where this relation has some universal properties \cite{Bueno:2015rda}. The existence and further generalizations of this universal ratio in more general cases is studied in several directions in  
\cite{{Bueno:2015xda},{Pang:2015lka},{Alishahiha:2015goa},{Miao:2015dua},{Bueno:2015qya},{Bueno:2015lza},{Bueno:2015ofa}}.

An important feature of the new divergence appearing in such entangling regions is the extensive contribution of its coefficient, $a(\Omega)$, to entanglement entropy which could be easily understood from the locality of the field theory.\footnote{Here locality means a one-to-one correspondence between Hilbert space decomposition and factorization of the spatial manifold of the QFT.} Since the UV contributions to entanglement entropy from separate points are supposed not to see each other, if there are more than one corner in the entangling region one may expect an extensive contribution to entanglement entropy from each corner (see \cite{Casini:2006hu}). We will come back to this point in this section and also in the body of this paper again, specifically when we study mutual information between sectors of an infinite circle.

%Although there is some universal information encoded in the EE, it has nothing to say about the field content of the QFT. 

As mentioned above, the EE for a single line segment in two dimensional CFTs is completely fixed by the central charge. In order to gain some information about the field content of the corresponding CFT, one needs to probe the theory by means of more powerful entanglement measures. In particular mutual information is such a measure which is defined for two disjoint entangling regions $A_1$ and $A_2$ as follows
\begin{align}\label{mutual}
I(A_1, A_2)=S_{A_1}+S_{A_2}-S_{A_1\cup A_2},
\end{align}
where $S_{A_1\cup A_2}$ is the entanglement entropy for the union of two entangling regions.

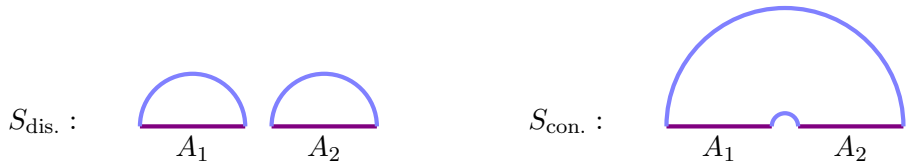
\begin{figure}
\begin{center}
\begin{tikzpicture}[scale=.7]
\draw[ultra thick,violet] (0,0) -- (2,0);
\draw[ultra thick,violet] (2.5,0) -- (4.5,0);
\draw[ultra thick,blue!50] (2,0) arc (0:180:1cm);
\draw[ultra thick,blue!50] (4.5,0) arc (0:180:1cm);
\draw[ultra thick,violet] (10,0) -- (12,0);
\draw[ultra thick,violet] (12.5,0) -- (14.5,0);
\draw[ultra thick,blue!50] (14.5,0) arc (0:180:2.25cm);
\draw[ultra thick,blue!50] (12.5,0) arc (0:180:0.25cm);
%\draw[step=1cm,gray,very thin] (0,-10) grid (25,25);
\draw[] (-1,0.1) node[left] {$S_{\text{dis.}}:$}; 
\draw[] (9,0.1) node[left] {$S_{\text{con.}}:$}; 
\draw[] (1,0) node[below] {$A_1$}; 
\draw[] (3.5,0) node[below] {$A_2$}; 
\draw[] (11,0) node[below] {$A_1$}; 
\draw[] (13.5,0) node[below] {$A_2$}; 
\end{tikzpicture}
\caption{Two different configurations for computing $S(A_1\cup A_2)$ using RT prescription.}
\end{center}
\label{fig:HMIconfig}
\end{figure}

Mutual information is a finite and positive quantity for specific entangling regions which measures the amount of entanglement shared between $A_1$ and $A_2$ regions. It is important to mention that mutual information  diverges when the separation between disjoint regions vanishes and they share a boundary. In reference \cite{Calabrese:2009ez} it was shown that MI is not only a function of the central charge of the corresponding two dimensional CFT rather it depends on the full operator content of the theory. 

On the other hand in the context of holographic CFTs, using the RT prescription it has been shown that holographic mutual information (HMI) exhibits a first order phase transition due to a discontinuity in its first derivative \cite{Headrick:2010zt}. It is believed that this phase transition is a reminiscent of the large central charge limit of the CFT and it disappears if one considers quantum corrections \cite{Faulkner:2013ana}. The gravity picture of this phase transition is simply due to a jump between two different configurations candidate for the minimal surface of the entanglement entropy of the union region $S_{A_1\cup A_2}$ (see Fig.\ref{fig:HMIconfig}).
This figure demonstrates two possible configurations corresponding to the HEE for the union of two entangling regions.
%An explicit computation shows that when the ratio between the length of the entangling regions and their separation exceeds a critical value, HMI has a finite value and otherwise it becomes zero.
HMI either vanishes or takes a finite value depending on the value of the ratio between the length of the entangling regions and their separation. It vanishes on one side of a critical value for this ratio and takes a finite value on the other side.
This can be understood as follows: when the disjoint regions are close enough together there is a finite correlation between them, but as they get far apart, the mutual correlation decreases and finally vanishes. 

%It is worth to note that the finiteness of mutual information depends on both the geometry of the entangling regions and also the spatial dimension of the field theory. For instance although mutual information is a UV-finite quantity for infinite strips in a four dimensional field theory it is not UV-finite for spherical entangling regions. 

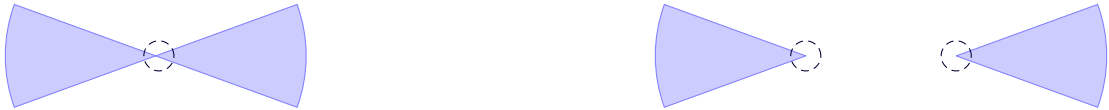
\begin{figure}
\begin{center}

\begin{tikzpicture}[scale=20]
  \fill[fill=blue!20!white, draw=blue!50!white]
    (0,0) -- (.94mm,-.342mm) arc (-20:20:1mm) -- (0,0);
%  \fill[fill=blue!20!white, draw=blue!50!white]
%    (-1mm,0) -- (-1.94mm,-.342mm) arc (160:120:1mm) -- (-1mm,0);
  \fill[fill=blue!20!white, draw=blue!50!white]
    (-1mm,0) -- (-1.94mm,.342mm) arc (160:200:1mm) -- (-1mm,0);
%%%%%%%%%%%%%%%%%%
%  \fill[fill=blue!20!white, draw=blue!50!white]
%    (-3mm,0) -- (-3.94mm,-.342mm) arc (-20:20:1mm) -- (-3mm,0);
  \fill[fill=blue!20!white,rotate=20, draw=blue!50!white]
    (-5mm,1.82mm) -- (-6mm,1.82mm) arc (180:140:1mm) -- (-5mm,1.82mm);
  \fill[fill=blue!20!white,rotate=-20, draw=blue!50!white]
    (-5mm,-1.82mm) -- (-4mm,-1.82mm) arc (0:40:1mm) -- (-5mm,-1.82mm);
%  \fill[fill=blue!20!white,rotate=-5, draw=blue!50!white]
%      (-3mm,0) -- (-4mm,0mm) arc (0:-170:1mm) -- (-3mm,0);
%\draw[dashed] (-7mm,0mm)--(7mm,0mm);
\draw[densely dashed,blue!20!black] (0mm,0mm) circle(.1mm);    
\draw[densely dashed,blue!20!black] (-1mm,0mm) circle(.1mm);    
\draw[densely dashed,blue!20!black] (-5.3mm,0mm) circle(.1mm);    
%%%%%%%%%%%%%%%%%%
\end{tikzpicture} 
\end{center}
\caption{Right: Two singular entangling regions which their singularities are far away from each other. The dashed circle denotes the local region responsible for the main UV-divergent part of the singularity. As long as these points are far from each other, which means their distance $d$ is much larger than the UV cut-off $1/\epsilon$, they contribute to the entanglement entropy (between the union of these regions and its complement) extensively. Left: As these two singular regions get closer to each other (and finally touch), their local regions responsible for the new UV-divergent term intersect with each other and thus their contribution to entanglement entropy is no more extensive.}
\label{fig:MIS}
\end{figure}

In this paper we are mainly considering three dimensional field theories which we expect the mutual information between infinite strips and disks to be UV-finite. For the case of singular entangling regions since an extra UV-divergent term appears in the entanglement entropy one should consider the relevant divergent terms precisely. For a configuration like that in the right panel of Fig.\ref{fig:MIS}, since the singular points are far away from each other one may expect mutual information to still be a UV-finite quantity because of the extensivity of corner contributions to entanglement entropy in local field theories \cite{Casini:2006hu} which was discussed before. This is not the case when the distance between these points vanishes like the left panel of Fig.\ref{fig:MIS}. In such a case the singular points share a local region and thus the extensivity of the corner divergent terms breaks down.\footnote{We thank Matthew Headrick for referring and specially thank Horacio Casini for an insightful discussion about this point.} The main part of this paper is an example of this latter case in section \ref{sec:HEM}. We also investigate an explicit example of the former case in section \ref{sec:SCC} to explicitly show the finiteness of mutual information even between singular regions.

Beside mutual information, similar quantities are also defined to deal with disjoint regions. Specifically for a system which is composed of at least three disjoint subsystems, another quantity which is called tripartite information is defined as \cite{Hayden:2011ag}
\begin{align}\label{tripartite0}
I^{[3]}(A_1,A_2,A_3)=S_{A_1}+S_{A_2}+S_{A_3}-S_{A_1\cup A_2}-S_{A_1\cup A_3}-S_{A_2\cup A_3}+S_{A_1\cup A_2\cup A_3},
\end{align}
where the $A_i$'s refer to disjoint entangling regions. Tripartite information for smooth subregions is known to be a UV-finite quantity. Remember that while we are dealing with two smooth subregions, if these regions share a boundary the mutual information between them is no longer a UV-finite quantity. In contrast with mutual information, tripartite information is known to be UV-finite even when the smooth regions share boundaries \cite{{Hayden:2011ag},{Allais:2011ys}}. Again we show in section \ref{sec:HEM} that this is not the case for singular surfaces when the shared boundary is the same as the singular point. This could be understood as a straightforward generalization of what was explained in Fig.\ref{fig:MIS}.

Another important property of tripartite information is the sign of this quantity. In contrast with mutual information, tripartite information in general may be either negative, positive or zero. In the holographic context it has been shown that holographic tripartite information has a definite sign and it is always negative. This property is also known as the monogamy property of HMI \cite{Hayden:2011ag}. 

The main goal of this paper is to investigate the holographic mutual and tripartite information and their possible phase transitions in the presence of a kink or a crease singularity in the entangling region. Indeed finding the minimal area surface corresponding to the union of entangling regions is a subtle task. This problem is already only solved analytically for parallel strips or concentric circles\footnote{These concentric circles can be mapped to two disjoint disks using a conformal transformation \cite{Fonda:2014cca}.}  as entangling regions \cite{{Headrick:2010zt},{Fischler:2012uv},{Fonda:2014cca},{Nakaguchi:2014pha}}. In particular the authors of \cite{Fonda:2014cca} have introduced an elegant numerical method using Surface Evolver to overcome this difficulty. By employing this numerical technique they could find HMI for various more complicated entangling regions\footnote{See \cite{Fonda:2015nma} where the generalization of this method to more general backgrounds is also considered.}. Of course here we do not employ this numerical method. We show that a simple observation can help us to find the HMI for sectors of a single infinite circle using the result of HEE for a kink entangling region in three dimensions. We also show that for more general configurations of this type one can calculate other entanglement measures e.g. holographic tripartite and $n$-partite information. 

The remainder of this paper is organized as follows: in section \ref{sec:rev} we briefly review the related literature on holographic entanglement entropy for singular surfaces. In section \ref{sec:HEM} we introduce our geometric set-up and study the holographic mutual, tripartite and $n$-partite information where we mainly focus on possible phase transitions of holographic mutual information. In section \ref{sec:HD} we study the possible generalizations of our set-up in section \ref{sec:HEM} from three dimensional CFTs to higher dimensions. In section \ref{sec:SCC} we study mutual information between singular regions without a shared boundary as an explicit example were the new divergent term is extensive, in contrast with what was studied in \ref{sec:HEM} and \ref{sec:HD} where we only study singular surfaces with a common singular point. In the last section beside a summary and some concluding remarks we also discuss about constructing a finite quantity from mutual information for singular surfaces and also discuss about the first law of entanglement for such configurations. Appendix \ref{sec:app1} is devoted to a comparison between singular surfaces in the sharp limit with the well-known results for strip entangling regions. In appendix \ref{sec:Hs} we give some results for the entanglement entropy of a kink in holographic theories with a hyperscaling violating geometry as their gravity dual.

\section{Holographic Entanglement Entropy (revisited)}\label{sec:rev}
In this section we shortly review the holographic entanglement entropy for singular surfaces which has been previously studied in \cite{{Hirata:2006jx},{Myers:2012vs}}. As mentioned in the previous section in a three dimensional theory only one type of singularity is possible which we refer to it by kink singularity (see the left panel of Fig.\ref{fig:kc}). Most of the analysis of this paper is devoted to this type of singularity. In higher dimensions various types of singularities are possible where two specific ones known as crease (see the right panel of Fig.\ref{fig:kc}) and cone have been studied previously. We will study crease entangling regions in section \ref{sec:HD}.\footnote{We do not consider conical entangling regions in this paper. The interested reader may refer to \cite{Myers:2012vs}.}

In order to clarify the definition of these surfaces consider a $d$ dimensional flat space time, i.e., $R^{1,d-1}$, as follows
\begin{align}\label{metric1}
ds^2=-dt^2+d\rho^2+\rho^2\left(d\theta^2+\sin^2\theta\; d\Omega_n^2\right)+\sum_{i=1}^{m}dx_i^2
\end{align}
where $d=n+m+3$ and $d\Omega_n$ corresponds to the metric of unit $n$-sphere. Following the terminology of \cite{Myers:2012vs} a kink is defined for $d=3$ and $m=n=0$ which is given by $k=\{t=0, 0<\rho<\infty, -\frac{\Omega}{2}\leq\theta \leq\frac{\Omega}{2}\}$.  A crease ($k\times R^{m}$) is the extension of kink to higher dimensions with $n=0$ and $d=3+m$. In this section we only consider kinks in three dimensions, i.e., $m=0$ so the bulk geometry is given by an AdS$_4$ space-time with the following metric
\begin{align}\label{metric}
ds^2=\frac{L^2}{z^2}\left(dz^2-dt^2+d\rho^2+\rho^2 d\theta^2\right),
\end{align}
where the spatial part of the boundary metric is considered in polar coordinates with $\rho$ and $\theta$ as the radial and azimuthal angle respectively. Also $z$ is the radial coordinate of the bulk geometry.

\begin{figure}
\begin{center}
\includegraphics[scale=2]{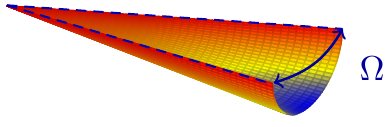}
\end{center}
\caption{Schematic representation of RT surface for a kink entangling region with opening angle $\Omega$ in a three dimensional CFT.}
\label{fig:schematicKink}
\end{figure}

The kink entangling region in three dimensions is defined as
\begin{equation}\label{entangling}
t=\mathrm{const.}\;\;\;,\;\;\;0<\rho< H\;\;\;,\;\;\;-\frac{\Omega}{2}\leq\theta \leq\frac{\Omega}{2},
\end{equation}
where $\Omega$ is the opening angle and $H$ is an IR cut-off on the radial coordinate. Due to the scaling symmetry on the bulk radial coordinate and the boundary coordinates, it was shown in \cite{Hirata:2006jx} that the RT surface in the bulk can be parametrized as $z(\rho,\theta)=\rho\;h(\theta)$ such that $h(\pm\frac{\Omega}{2})=0$. See Fig.\ref{fig:schematicKink} for a schematic plot of the minimal surface. In this case the HEE functional becomes 
\begin{align}\label{heefunc}
S=\frac{L^2}{2G_N}\int_{\frac{\epsilon}{h_*}}^H \frac{d\rho}{\rho}\int_0^{\frac{\Omega}{2}-\delta}d\theta\frac{\sqrt{1+h^2+h'^2}}{h^2},
\end{align}
where $\epsilon$ is the inverse UV cut-off defined by $\epsilon=\rho\;h(\frac{\Omega}{2}-\delta)$, and $h_*=h(0)$ is the turning point of the RT surface in the bulk which is defined where $h'(0)=0$ (prime denotes the derivative with respect to $\theta$).

Applying the standard variational principle leads to the surface which minimizes this functional. According to Eq.\eqref{heefunc} where the integrand does not depend explicitly on $\theta$, the corresponding Hamiltonian is a conserved quantity 
\begin{align}\label{thetamom}
\mathcal{H}\equiv \frac{1+h^2}{h^2\sqrt{1+h^2+h'^2}}=\frac{\sqrt{1+h_*^2}}{h_*^2}.
\end{align}
Using this relation it is an easy task to find the profile of the minimal surface and hence the HEE. The final result for the HEE is
\begin{align}\label{HEE}
S(\Omega)=\frac{L^2}{2G_N}\frac{H}{\epsilon}-a(\Omega)\log \frac{H}{\epsilon}-\left(\frac{\pi L^2}{4G_N h_*}+a(\Omega)\log h_*\right)+\mathcal{O}\left(\frac{\epsilon}{H}\right),
\end{align}
where the function $a(\Omega)$ is defined as
\begin{align}\label{aomega}
a(\Omega)=\frac{L^2}{2G_N}\int_0^\infty dy\left[1-\sqrt{\frac{1+h_*^2(1+y^2)}{2+h_*^2(1+y^2)}}\right],
\end{align}
and the boundary data (the opening angle $\Omega$) is given in terms of the bulk turning point $h_*$ as
\begin{align}\label{omehah0}
\Omega =\int_0^{h_*} dh\frac{2h^2\sqrt{1+h_*^2}}{\sqrt{1+h^2}\sqrt{h_*^4(1+h^2)-h^4(1+h_*^2)}}.
\end{align}

\begin{figure}
\begin{center}
\includegraphics[scale=.6]{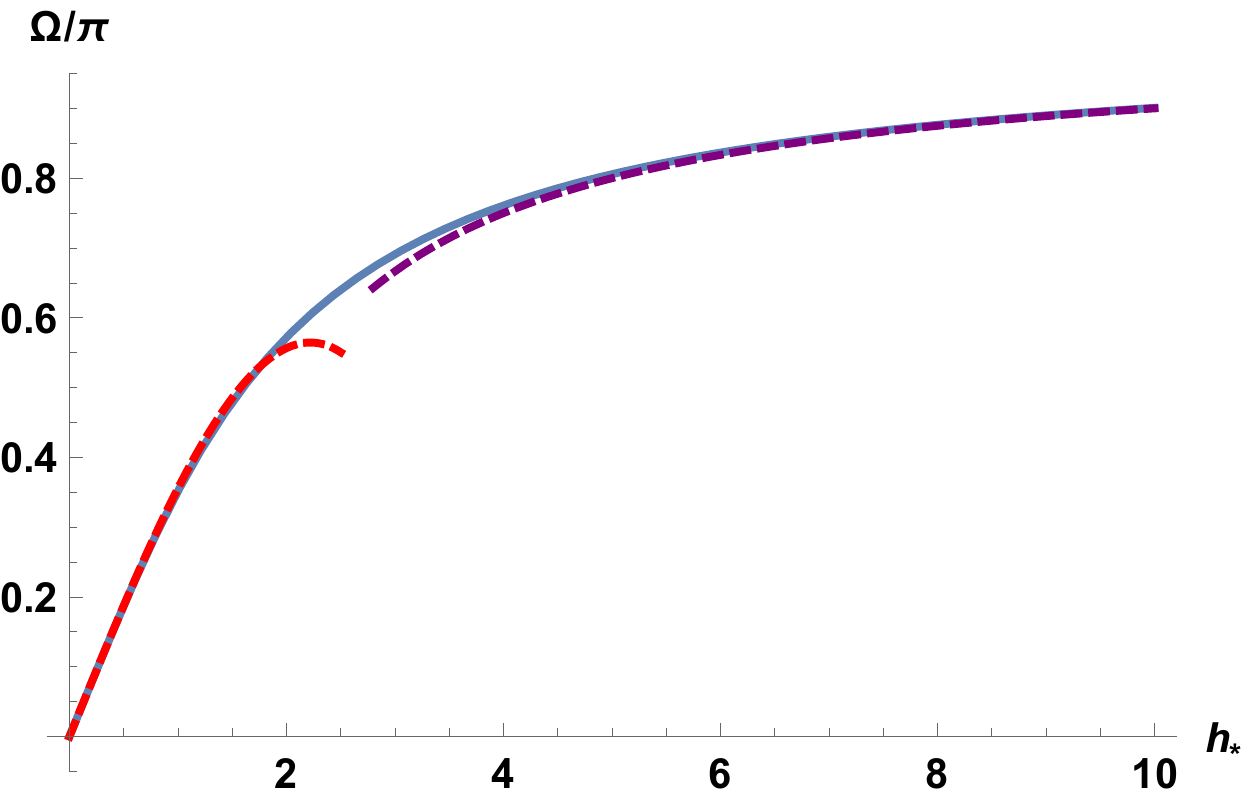}
\includegraphics[scale=.6]{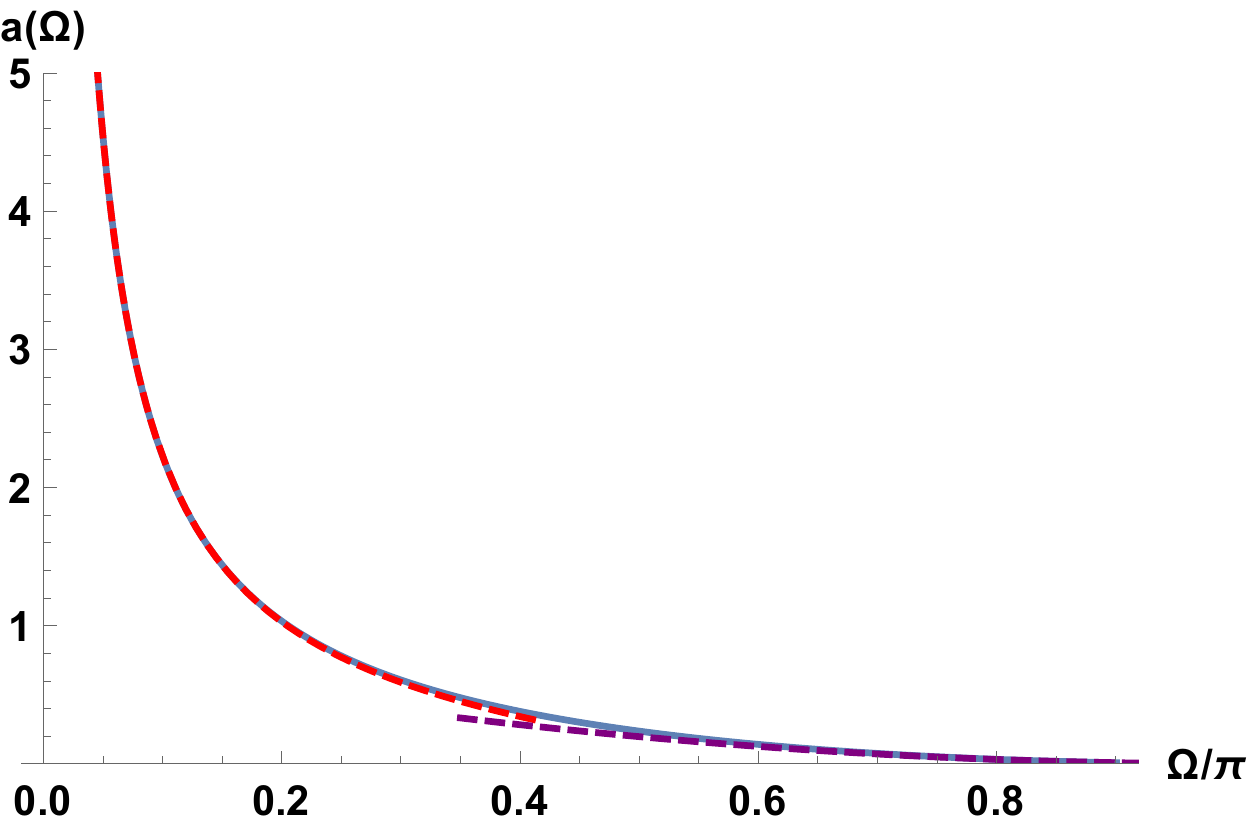}
\end{center}
\caption{\textit{Left}: $\Omega/\pi$ as  a function of the turning point $h_*$. \textit{Right}: $a$ as a function of the opening angle $\Omega$. In both plots the red and violet curves correspond to the small opening angle and smooth limit, respectively.}
\label{fig:aomega}
\end{figure}
It is interesting to focus on two specific limits where the kink is extremely sharp $(\Omega \rightarrow 0)$ or it is a smooth surface which is slightly folded $(\Omega \rightarrow \pi)$. In these two limits one can work out the HEE analytically as follows \cite{Bueno:2015xda}:

\vspace{4mm}
1) Sharp limit $(\Omega \rightarrow 0)$
\begin{align}\label{smallregion}
\begin{split}
\Omega &=\frac{2\sqrt{\pi}\Gamma(3/4)}{\Gamma(1/4)}h_*-\frac{[3\Gamma^2(3/4)-\Gamma(1/4)\Gamma(5/4)]}{6\sqrt{2\pi}}h_*^3+\cdots,\\
a(\Omega)&=\frac{\kappa}{\Omega}-\frac{L^2}{G_N}\frac{\pi \Gamma(1/4)}{48\sqrt{2}\Gamma^3(3/4)}\Omega+\cdots,\\
S(\Omega)&=\frac{L^2}{2G_N}\frac{H}{\epsilon}-\left[\frac{L^2}{2G_N}\frac{\pi^{3/2}\Gamma(3/4)}{\Gamma(1/4)}+\kappa \left(\log \frac{H}{\epsilon}+\log \Omega+\log\frac{2\Gamma(5/4)}{\sqrt{\pi}\Gamma(3/4)}\right)\right]\frac{1}{\Omega}+\cdots\\
&\sim \frac{L^2}{2G_N}\frac{H}{\epsilon}-\frac{\kappa}{\Omega}\left(\log \frac{H}{\epsilon}+\log \Omega\right)+\cdots,
\end{split}
\end{align}
where $\kappa=\frac{L^2}{2\pi G_N}\Gamma^4(3/4)$. Note that assuming $\Omega\ll\frac{H}{\epsilon}$ we will neglect the last term in the above expression for the HEE in the following discussion. 

\vspace{4mm}
2) Smooth limit $(\Omega \rightarrow \pi)$
\begin{align}\label{largeregion}
\begin{split}
\Omega &=\pi-\frac{\pi}{h_*}+\cdots\\
a(\Omega)&=\sigma(\pi-\Omega)^2+\cdots,\\
S(\Omega)&=\frac{L^2}{2G_N}\frac{H}{\epsilon}-\frac{L^2}{2G_N}\frac{\pi-\Omega}{2}-\sigma \left(\log \frac{H}{\epsilon}-\log \left(1-\frac{\Omega}{\pi}\right)\right)(\pi-\Omega)^2+\cdots,
\end{split}
\end{align}
where $\sigma=\frac{L^2}{8\pi G_N}$.

Fig.\ref{fig:aomega} demonstrates the behavior of $\Omega(h_*)$ and $a(\Omega)$. In this figure we also included the above asymptotic results which coincide with the exact results in a wide range of opening angles. 
%In Fig.\ref{fig:HEE} we have plotted the area substracted entanglement entropy which is defined as
%\begin{align}
%\tilde{S}(\Omega)=S(\Omega)-\frac{L^2}{2G_N}\frac{H}{\epsilon}.
%\end{align}
Note that in this figure we have set $\frac{L^2}{2G_N}=1$. 

%Also note that in the smooth limit $(\Omega \rightarrow \pi)$, the cut-off dependence disappears and the value of the entanglement entropy for different values of $\epsilon$ coincide. This is expected for smooth entangling regions and it is also understood from the expression of HEE in Eq.\eqref{largeregion}, where at the first order of $\tilde{S}(\Omega)$ does not depend on the UV cut-off.

%\begin{figure}
%\begin{center}
%\includegraphics[scale=.8]{HEEomega.pdf}
%\end{center}
%\caption{Area substracted Entropy $\tilde{S}(\Omega)$ in unites where $H=1$ for different values of the UV cut-off. Note that in the smooth limit these curves become independent of $\epsilon$ and coincide.}
%\label{fig:HEE}
%\end{figure}

\section{Holographic Entanglement Measures ($d=3$)}\label{sec:HEM}
In this section we aim to study holographic entanglement measures with our main focus on holographic mutual information between kink entangling regions in three dimensions. As mentioned in the introduction section the main subtlety to compute quantities such as mutual information is how to compute the last term in Eq.\eqref{mutual}. The core of our idea is to consider the set of kinks in interest as sectors of an infinite circle (see Fig.\ref{fig:I} and Fig.\ref{fig:I3}). In such a configuration the holographic entanglement entropy of a union of two kink entangling regions can be expressed in terms of minimal surfaces anchoring to certain kinks on the boundary. See Fig.\ref{fig:regionsmutual} and Fig.\ref{fig:I3config} for schematic graphical visualizations. 
\subsection{Holographic Mutual Information}
In this subsection we study the holographic mutual information for the specific configuration mentioned above. The entangling regions are sectors of a single infinite circle with opening angles $\Omega_1$ and $\Omega_2$. These entangling regions touch each other at a single point which is the center of the infinite circle. We denote the angular separation between these regions with $\omega$ (see Fig. \ref{fig:I}). Note that for a bipartite entangling region there are always two sectors in-between $\Omega_1$ and $\Omega_2$ sectors which we define $\omega$ to be the minimum value of the opening angle between these two sectors. By this definition the value of $\omega$ is restricted to $0\le \omega \le \pi-\left(\Omega_1+\Omega_2\right)/2$. We will consider regions with no angular overlap thus we will always assume $\Omega_1+\Omega_2+\omega < 2\pi$. 
\begin{figure}
\begin{center}
\begin{tikzpicture}[scale=6]
  \fill[fill=blue!20!white,rotate=-15, draw=blue!50!white]
    (0,0) -- (3mm,0mm) arc (0:60:3mm) -- (0,0);
  \fill[fill=blue!20!white,rotate=75, draw=blue!50!white]
    (0,0) -- (3mm,0mm) arc (0:45:3mm) -- (0,0);
        \draw[thick,dashed,blue!40!black] (0cm,0cm) circle(3mm);    
%%%%%%%%%%%%%%%%%%
  \draw[blue!40!black,<->, rotate=-15]
    (3.5mm,0mm) arc (0:60:3.5mm);
  \draw[red!60,thick,<->, rotate=45]
    (3.5mm,0mm) arc (0:30:3.5mm);
  \draw[blue!40!black,<->, rotate=75]
    (3.5mm,0mm) arc (0:45:3.5mm);
\draw [blue!40!black] (4mm,0.1) node {$\Omega_1$};
\draw [red!60] (1.9mm,3.4mm) node {$\omega$};
\draw [blue!40!black] (-0.6mm,4mm) node {$\Omega_2$};
\end{tikzpicture} 
\end{center}
\caption{The configuration of a bipartite entangling region. $\Omega_i$'s are the opening angles and $\omega$ is the angular separation and the radial coordinate which runs over $0\le \rho<\infty$.}
\label{fig:I}
\end{figure}
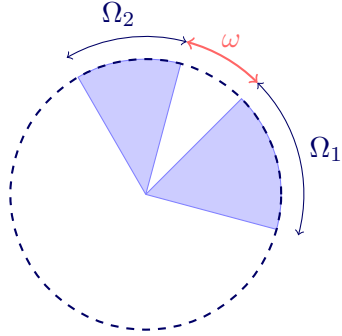
\begin{figure}
\begin{center}
\includegraphics[scale=1.3]{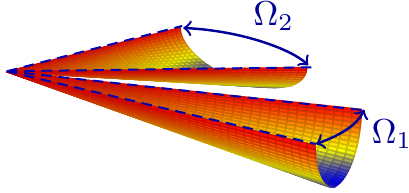}
\includegraphics[scale=1.3]{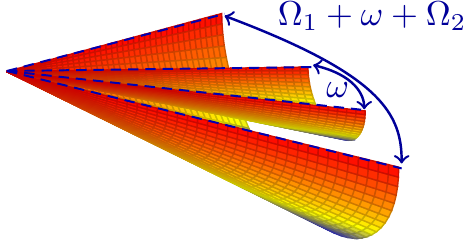}
\end{center}
\caption{Schematic representation of the RT surfaces corresponding to $S_{\Omega_1 \cup \Omega_2}$ for disconnected (left) and connected (right) configurations.}
\label{fig:regionsmutual}
\end{figure}

Considering such entangling regions, from Eq.\eqref{mutual} one finds the mutual information between these sectors as
\begin{align}\label{mutual1}
I(\Omega_1, \Omega_2)=S_{\Omega_1}+S_{\Omega_2}-S_{\Omega_1 \cup \Omega_2}.
\end{align}
The first two terms in the above equation are easy to compute and the main challenge is to find the suitable expression for the last one, i.e., $S_{\Omega_1 \cup \Omega_2}$. Indeed there are two different configurations which are competing for the value of this term. These two configurations which we denote them by ``connected" and ``disconnected" configurations are shown schematically in Fig.\ref{fig:regionsmutual}. According to this figure the value of $S_{\Omega_1 \cup \Omega_2}$ is given by one of the following expressions
\begin{align}\label{SAUB}
S_{\Omega_1 \cup \Omega_2}=
\begin{cases}
S_{\Omega_1+\Omega_2+\omega}+S_{\omega}\left(\equiv S_{\text{con.}}\right)  & ~~ \omega\ll \{\Omega_1,\Omega_2\}\\
S_{\Omega_1}+S_{\Omega_2}\left(\equiv S_{\text{dis.}}\right) & ~~ \omega\gg \{\Omega_1,\Omega_2\} 
\end{cases}
\end{align}
depending on the value of three parameters $\Omega_1$, $\Omega_2$, and $\omega$. Note that the contribution of the disconnected configuration is independent of the value of $\omega$. As we will see our numerical results show that there is always a transition between connected and disconnected configurations varying the values of these parameters.

In order to simplify our calculations, we restrict the following discussion to the case of equal opening angles, i.e., $\Omega_1=\Omega_2=\Omega$. In this case we can find a simple explicit expression for HMI which is given by
\begin{align}\label{equalregionsMI}
I&=2\tilde{S}_\Omega-\min\{2\tilde{S}_\Omega, \tilde{S}_{2\Omega+\omega}+\tilde{S}_\omega\},\nonumber\\
\tilde{S}_\Omega &=-a(\Omega) \log \frac{H}{\epsilon}-\left(\frac{\pi L^2}{4G_N h_*(\Omega)}+a(\Omega)\log h_*(\Omega)\right),\nonumber\\
\tilde{S}_\omega &=-a(\omega) \log \frac{H}{\epsilon}-\left(\frac{\pi L^2}{4G_N h_*(\omega)}+a(\omega)\log h_*(\omega)\right),\nonumber\\ \tilde{S}_{2\Omega+\omega}&=-a(2\Omega+\omega)\log \frac{H}{\epsilon}-\left(\frac{\pi L^2}{4G_N h_*(2\Omega+\omega)}+a(2\Omega+\omega)\log h_*(2\Omega+\omega)\right).
\end{align}

\begin{figure}
\begin{center}
\includegraphics[scale=.25]{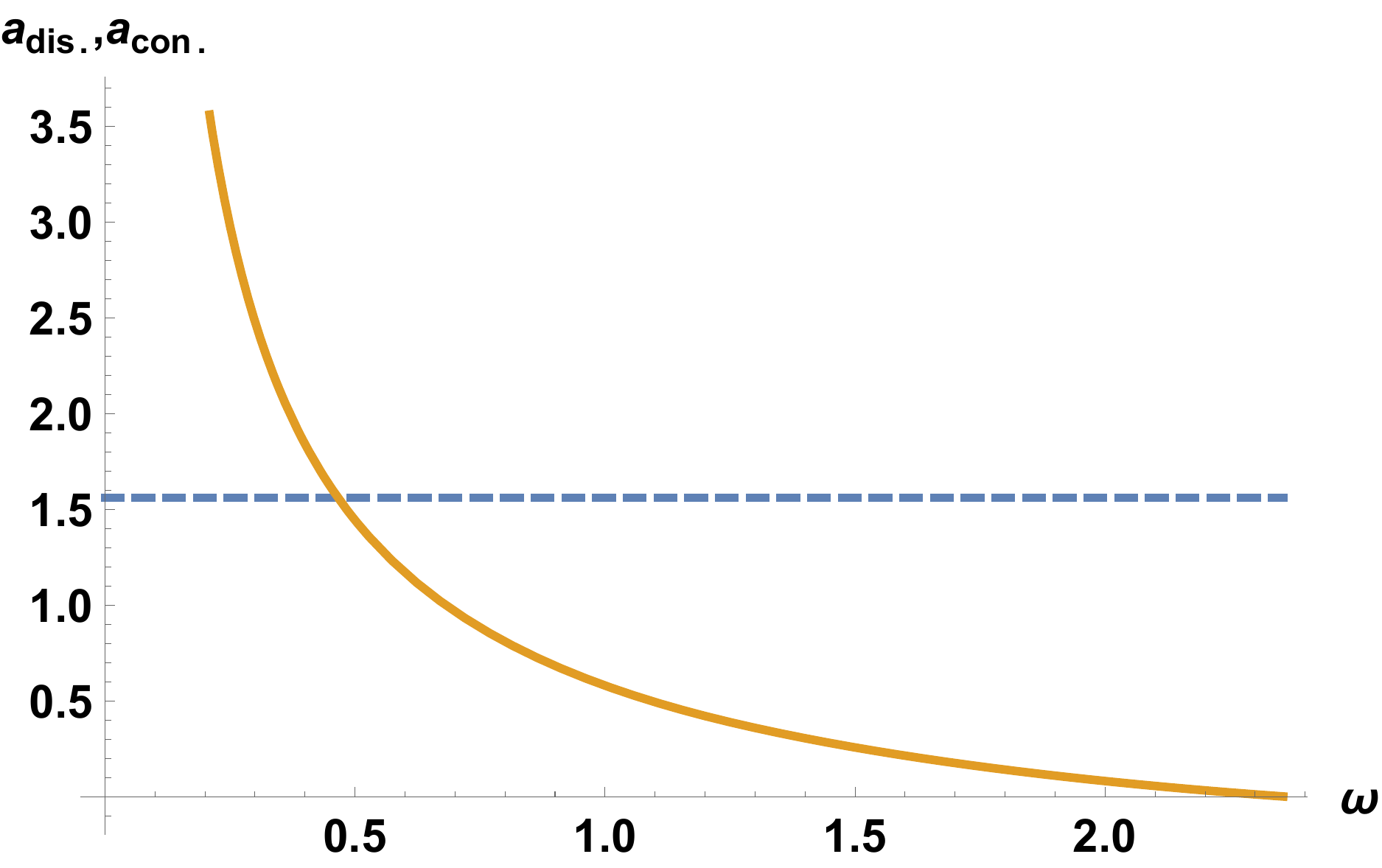}
\includegraphics[scale=.25]{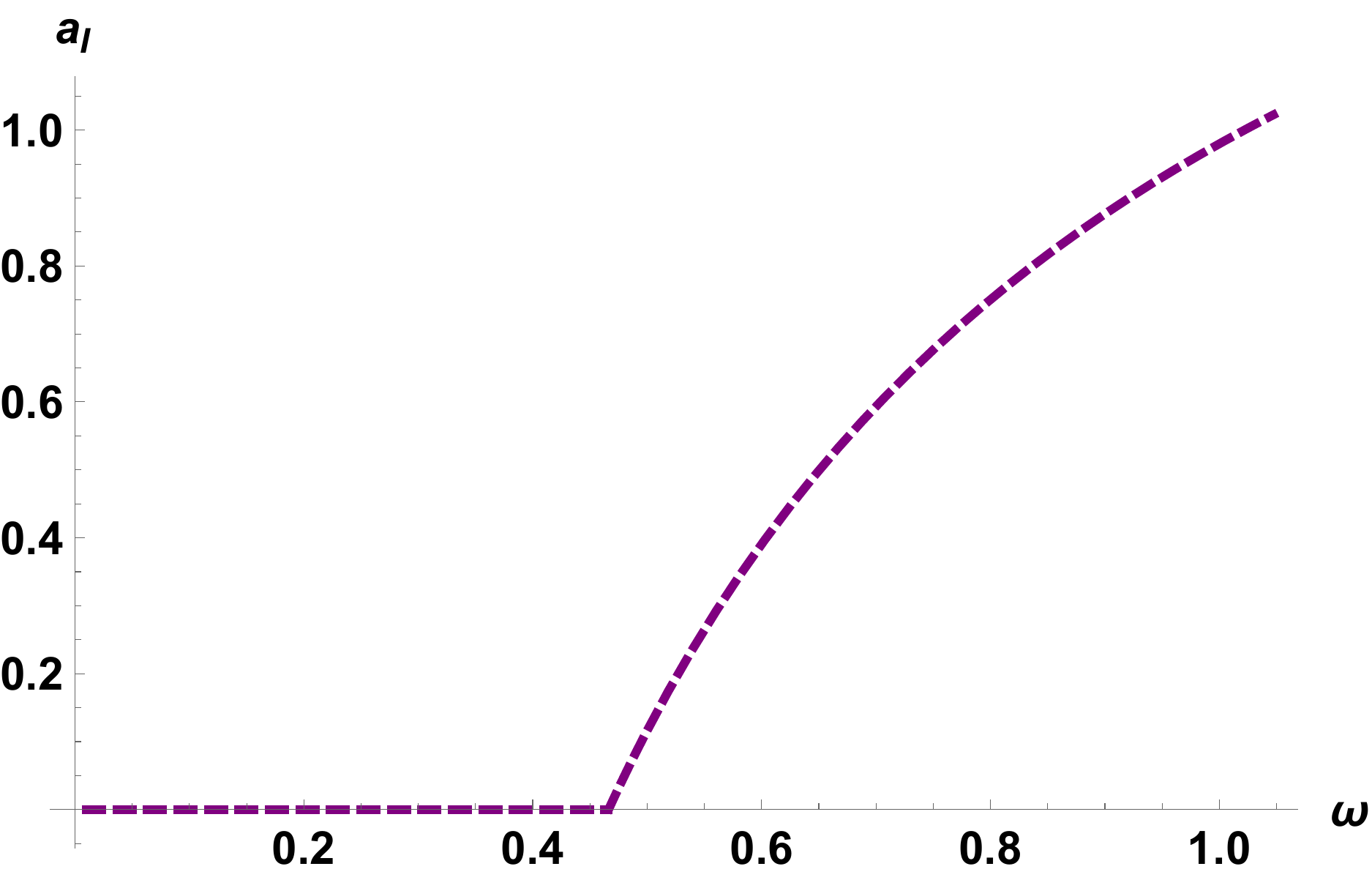}
\includegraphics[scale=.25]{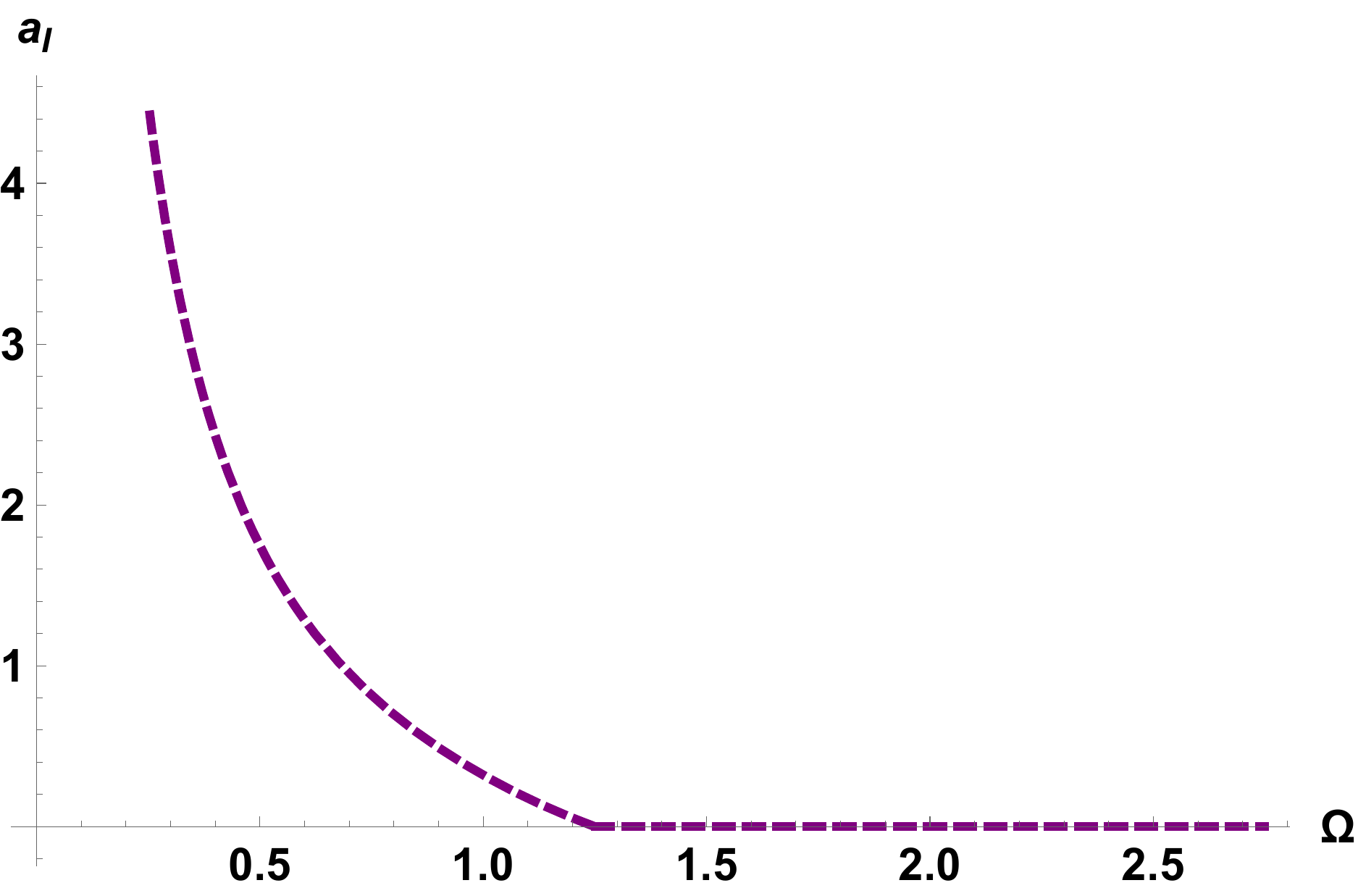}
\end{center}
\caption{\textit{Left}: Comparing the contributions of connected (solid curve) and disconnected (dashed curve) configurations to the universal part of HEE with $\Omega=\frac{\pi}{4}$. \textit{Middle}: The coefficient of the universal part of HMI as a function of $\omega$ for $\Omega=\frac{\pi}{4}$.  \textit{Right}: The coefficient of the universal part of  HMI as a function of $\Omega$ for $\omega=\frac{\pi}{4}$.}
\label{fig:mutual}
\end{figure}

\begin{figure}
\begin{center}
\includegraphics[scale=.6]{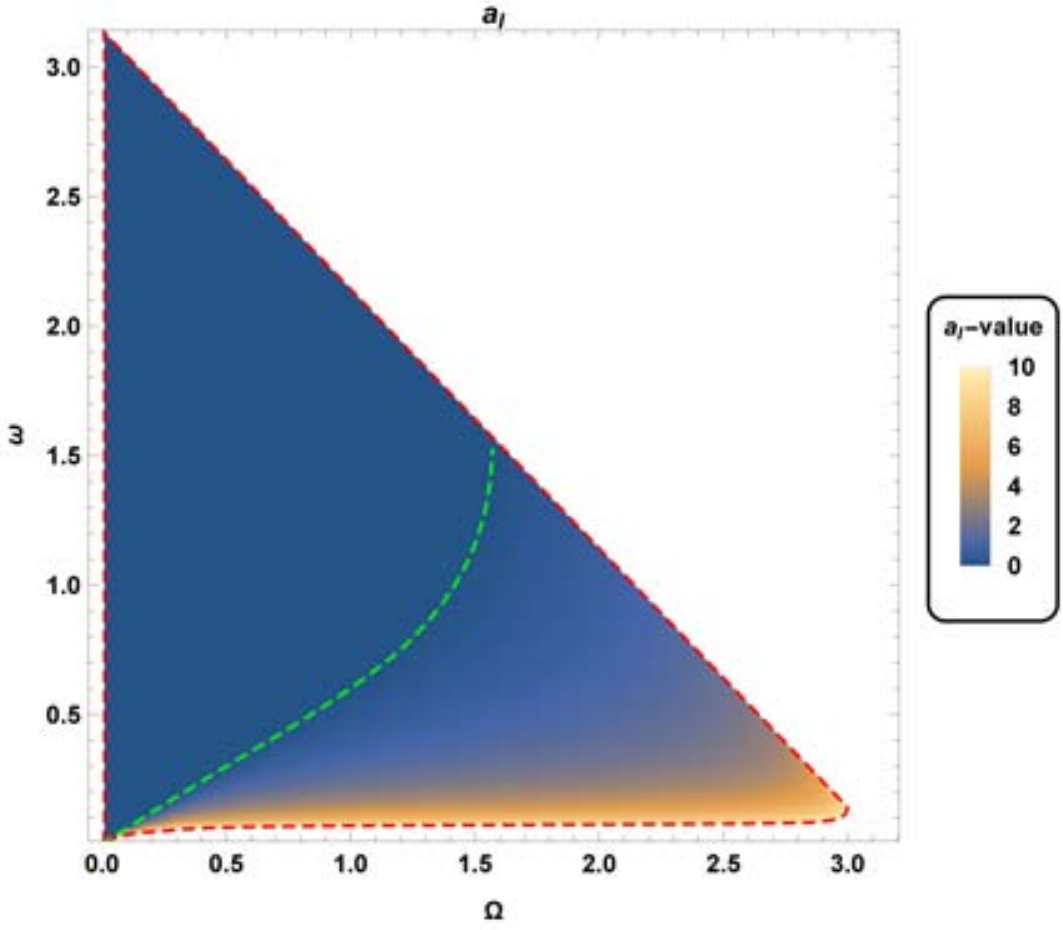}
%\hspace{5mm}
%\includegraphics[scale=.35]{Iepsilon.pdf}
%\includegraphics[scale=.7]{densityplot-I3}
\end{center}
\caption{Density plot of the universal part of HMI for regions with equal opening angles. The dashed green curve which is simply a guide to the eye is the transition curve and the dashed red boundary is $2\Omega+\omega <2\pi$ constraint. The white region below this boundary corresponds to the divergence of HMI in the  small separation limit. }
\label{fig:densityplot-I3}
\end{figure}
In Fig.\ref{fig:mutual} we have summarized the behavior of the universal part of HMI in this simple set-up. In the left plot the contributions of connected and disconnected configurations to the universal part of HEE for a fixed opening angle are compared. These quantities are defined as follows
\begin{align}
a_{\rm dis.}=2a_{\Omega},\;\;\;a_{\rm con.}=a_{2\Omega+\omega}+a_{\omega}.
\end{align}
%Although the \textit{existence} of the transition is universal, these plots show that the transition point is not universal since it depends on the UV cut-off.
In the middle and right plots we have demonstrated the universal part of HMI which is defined as 
\begin{align}\label{aI}
a_{I}=-a_{\rm dis.}-\min \{-a_{\rm dis.},-a_{\rm con.}\}
\end{align}
as a function of $\omega$ and $\Omega$ respectively. According to these plots by fixing the opening angles and increasing the separation between the regions the HMI vanishes. Also Fig.\ref{fig:densityplot-I3} is a density plot which shows the behavior and transition points of the universal part of HMI. It is not hard to show that the area divergent contributions to HMI cancel out and the remaining $\epsilon$-dependence is logarithmic.
%\begin{figure}
%\begin{center}
%\includegraphics[scale=.8]{I-epsilon}
%\end{center}
%\caption{HMI as a function of UV cut-off for $\Omega=\frac{\pi}{2}$ and $\omega=\frac{\pi}{4}$ (blue curve). The dashed red curve produced using a logarithmic fit function.}
%\label{fig:I-epsilon}
%\end{figure}

\begin{figure}
\begin{center}
\includegraphics[scale=.8]{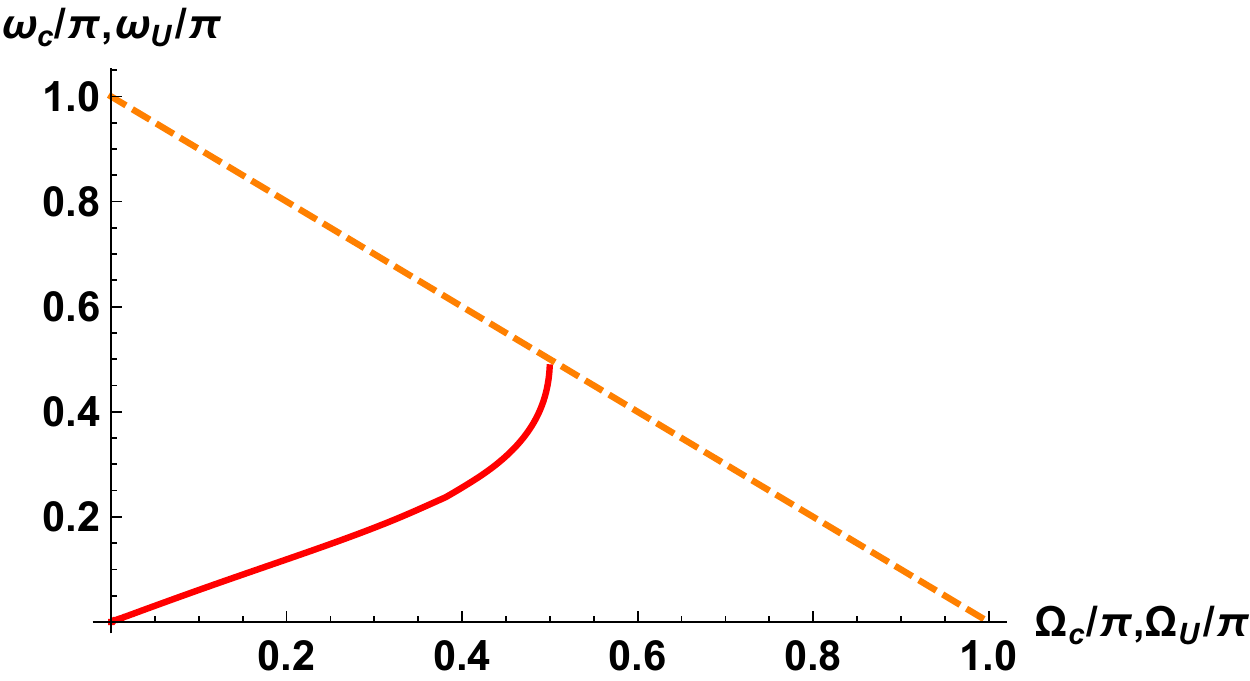}
\end{center}
\caption{The solid red curve shows the universal transition curve for HMI. HMI is non zero (positive) below the universal transition curve which is bounded from the other side by the boundary of the geometrical constraint $\omega+\Omega\le\pi$ shown by the dashed orange curve.
}
\label{fig:mutual-trans}
\end{figure}

Now we can further investigate the transition curve of the HMI in the parameter space. Such a curve is supposed to satisfy
\begin{align}\label{eq:a1}
2S_{\Omega_c}=S_{2\Omega_c+\omega_c}+S_{\omega_c},
\end{align}
which depends on $\epsilon$. One can easily check that as $\epsilon\to 0$, solutions of Eq.\eqref{eq:a1} merge to a cut-off independent curve which we call the universal transition curve for HMI. This universal curve is the solution of
\begin{align}\label{eq:a2}
2a_{\Omega_U}=a_{2\Omega_U+\omega_U}+a_{\omega_U},
\end{align}
where the logarithmic UV divergence of HMI cancels out. The universal transition curve is shown by a solid red curve in Fig.\ref{fig:mutual-trans}, where the dotted orange line corresponds to the boundary of the $\omega+\Omega\le \pi$ constraint which was mentioned previously.

In order to further explore the HMI for such configurations, we focus on the sharp kink and the slightly folded limits where we are able to give analytical expressions for HEE and thus HMI.

%\vspace{4mm}
\noindent\textbf{(i) Sharp limit} $\{\omega,\;\Omega\}\ll 1$:\\
In this case combining Eq.\eqref{smallregion} and Eq.\eqref{equalregionsMI} leads to 
\begin{align}\begin{split}\label{Ismall}
I&=-\frac{2\kappa}{\Omega}\log\frac{H}{\epsilon}+\min\left\{\frac{2}{\Omega},\frac{1}{2\Omega+\omega}+\frac{1}{\omega}\right\}\kappa \log\frac{H}{\epsilon}\\
&=
\begin{cases}
\left(-\frac{2}{\Omega}+\frac{1}{2\Omega+\omega}+\frac{1}{\omega}\right)\kappa \log\frac{H}{\epsilon}\sim\frac{\kappa}{\omega} \log\frac{H}{\epsilon} & ~~ \omega\ll \Omega \ll 1\\
0 & ~~ \Omega\ll \omega \ll 1.
\end{cases}
%&=\Bigg\{ \begin{array}{rcl}
%&\left(-\frac{2}{\Omega}+\frac{1}{2\Omega+\omega}+\frac{1}{\omega}\right)\kappa \log\frac{H}{\epsilon}\sim\frac{\kappa}{\omega} \log\frac{H}{\epsilon}&\,\,\,\,\,\omega\ll \Omega \ll 1,\\
%&0&\,\,\,\,\,\Omega\ll \omega \ll 1.
%\end{array}\,\,
\end{split}
\end{align}
Actually this final result can be understood considering a conformal map relating the corner geometry to a strip (see appendix A of \cite{Myers:2012vs}). 
This conformal map is given by
\begin{equation}\label{cmap}
t=L e^{\frac{Y}{L}}\cos \xi\;\;\;,\;\;\;\rho=L e^{\frac{Y}{L}}\sin \xi,
\end{equation}
and the entangling region \eqref{entangling} in the new geometry becomes
\begin{equation}
\xi=\frac{\pi}{2}\;\;\;,\;\;\;Y_-<Y<Y_+\;\;\;,\;\;\;-\frac{\Omega}{2}\leq\theta \leq\frac{\Omega}{2},
\end{equation}
where $Y_+=L\log \frac{H}{L}$ and $Y_-=L\log \frac{\epsilon}{L}$. Now by making use of the HEE for a strip the mutual information for two parallel strips with width $\ell$ and separation $x$ is given by \cite{{Fischler:2012uv},{Alishahiha:2014jxa}}
\begin{align}
I=
\begin{cases}
\kappa(Y_+-Y_-)\left(-\frac{1}{\ell}+\frac{1}{2(2\ell+x)}+\frac{1}{2x}\right) & ~~ x\ll \ell \\
0 & ~~ x \gg \ell
\end{cases},
%\Bigg\{ \begin{array}{rcl}
%&\kappa(Y_+-Y_-)\left(-\frac{1}{\ell}+\frac{1}{2(2\ell+x)}+\frac{1}{2x}\right)&\,\,\,\,\,x\ll \ell ,\\
%&0&\,\,\,\,\,x \gg \ell.
%\end{array}\,\,
\end{align}
where $Y_{\pm}$ are the regulator scales for the length of the strip. In the small opening angle limit , i.e., $\{\Omega, \omega\} \ll 1$,  one finds $\ell\equiv \Omega L \ll L$ and $x\equiv \omega L \ll L$ and the above expression reduces to Eq.\eqref{Ismall}. Also note that in $x\ll \ell$ limit only the last term contributes.

\begin{figure}
\begin{center}
\includegraphics[scale=.55]{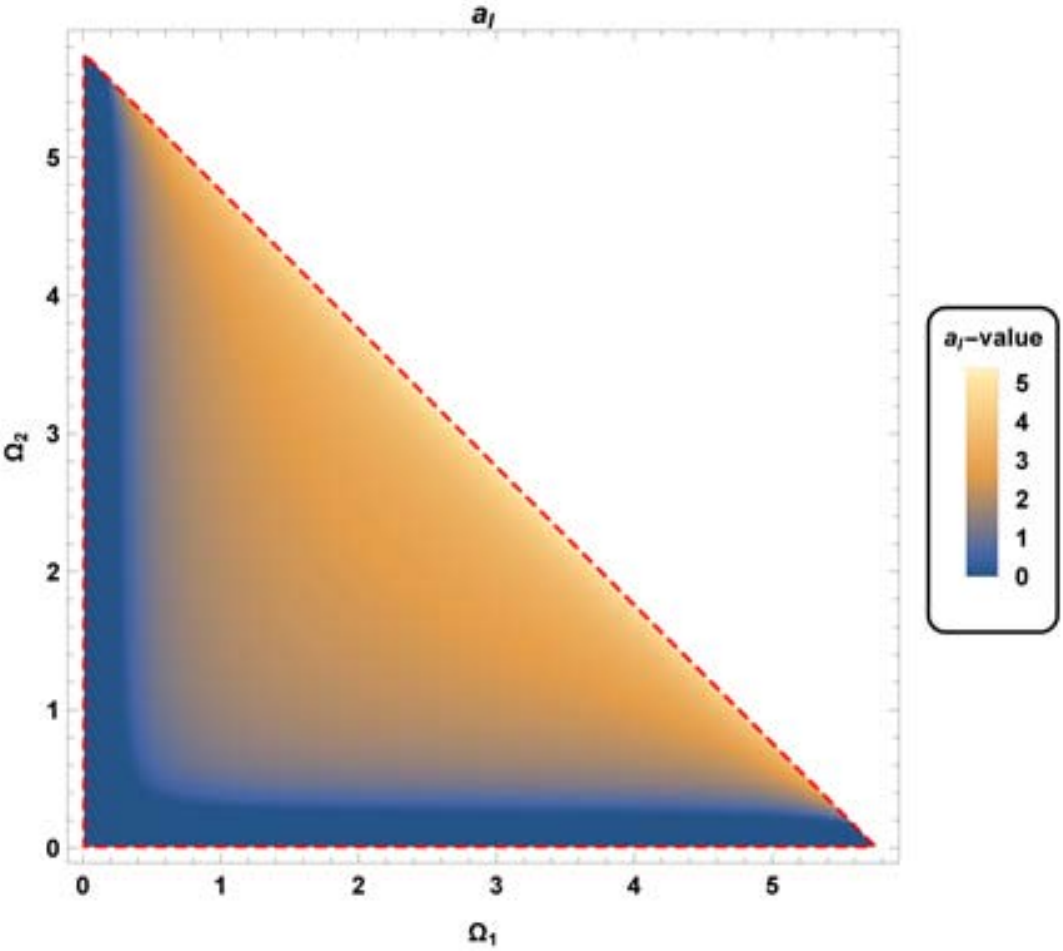}
\hspace{7mm}
\includegraphics[scale=.55]{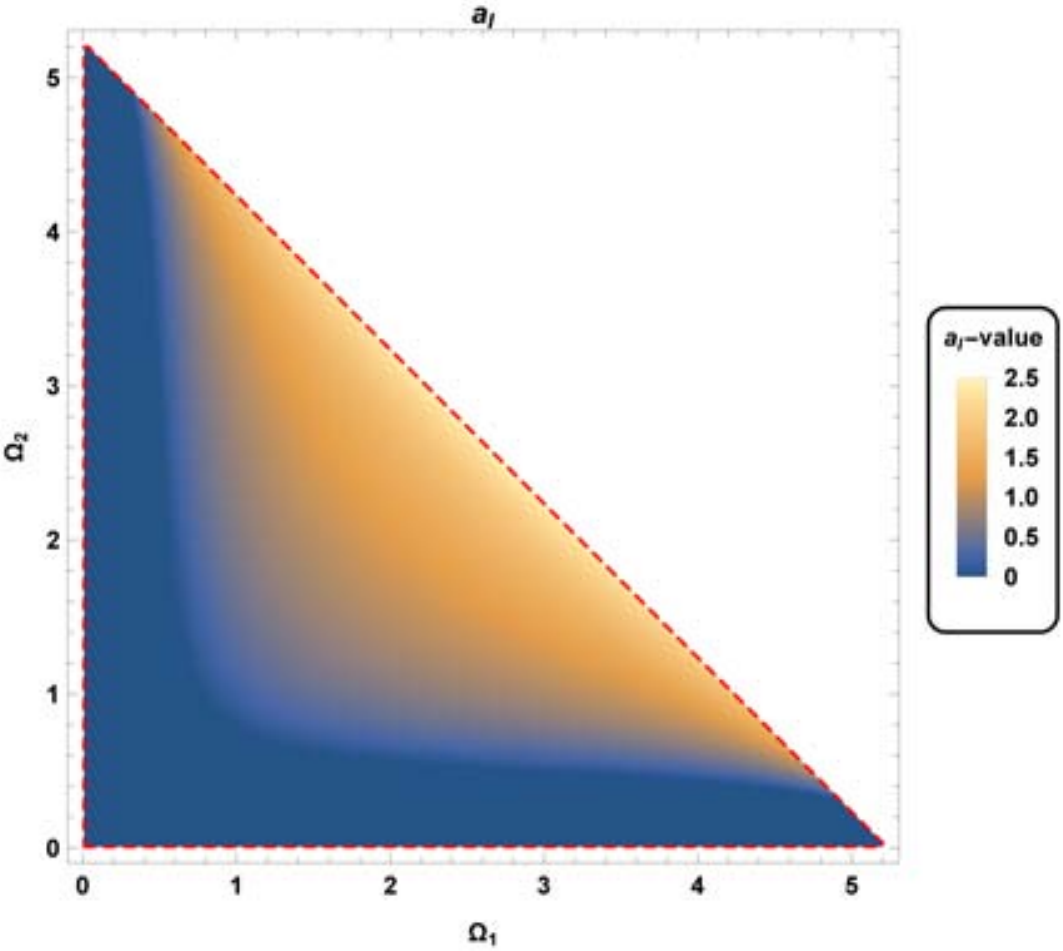}
\end{center}
\caption{Density plot of the universal part of HMI for $\omega=\frac{\pi}{12}$ (left) and $\omega=\frac{\pi}{6}$ (right). The dashed red curve is the boundary of the $\Omega_1+\Omega_2+2\omega< 2\pi$ constraint.}
\label{fig:densityplot-I1}
\end{figure}

%\vspace{4mm}
\noindent\textbf{(ii) Smooth limit} $\omega \sim 0,\; \Omega \sim \pi,\;\;\;\Omega+\omega=\pi$:\\
In this case the fact that we always consider a pure state requires that $S_{2\Omega+\omega}=S_{2\pi-\omega}=S_{\omega}$ so Eq.\eqref{equalregionsMI} becomes 
\begin{align}
I=2S_\Omega-2\min\{S_\Omega, S_\omega\}.
\end{align}
Using Eq.\eqref{largeregion} and Eq.\eqref{smallregion} for $S_\Omega$ and $S_\omega$ respectively the HMI at the leading order reduces to
\begin{align}\label{Ilarge}
I&=-\frac{L^2}{2G_N}\omega-\min\left\{-\frac{L^2}{2G_N}\omega,-\frac{2\kappa}{\omega}\log\frac{H}{\epsilon}\right\}\sim \frac{2\kappa}{\omega}\log\frac{H}{\epsilon}.
\end{align}

In order to investigate the behavior of HMI in this singular configuration more generally, we consider the case where the opening angles of the entangling regions are not equal, i.e., $\Omega_1\neq \Omega_2$. Fig.\ref{fig:densityplot-I1} demonstrates the density plot of the universal part of HMI for different values of $\omega$ as a function of entangling opening angles. These plots show the transition points of this quantity and also regions with non-vanishing HMI. 

\subsection{Holographic Tripartite Information}
\begin{figure}
\begin{center}
\begin{tikzpicture}[scale=6]
  \fill[fill=blue!20!white,rotate=-15, draw=blue!50!white]
    (0,0) -- (3mm,0mm) arc (0:60:3mm) -- (0,0);
  \fill[fill=blue!20!white,rotate=75, draw=blue!50!white]
     (0,0) -- (3mm,0mm) arc (0:45:3mm) -- (0,0);
   \fill[fill=blue!20!white,rotate=160, draw=blue!50!white]
    (0,0) -- (3mm,0mm) arc (0:35:3mm) -- (0,0);
        \draw[thick,dashed,blue!40!black] (0cm,0cm) circle(3mm);    
%%%%%%%%%%%%%%%%%%
  \draw[blue!40!black,<->, rotate=-15]
    (3.5mm,0mm) arc (0:60:3.5mm);
  \draw[red!60,thick,<->, rotate=45]
    (3.5mm,0mm) arc (0:30:3.5mm);
  \draw[red!60,thick,<->, rotate=120]
    (3.5mm,0mm) arc (0:40:3.5mm);
  \draw[blue!40!black,<->, rotate=75]
    (3.5mm,0mm) arc (0:45:3.5mm);
      \draw[blue!40!black,<->, rotate=160]
    (3.5mm,0mm) arc (0:35:3.5mm);
\draw [blue!40!black] (4mm,0.1) node {$\Omega_1$};
\draw [red!60] (1.9mm,3.4mm) node {$\omega_1$};
\draw [blue!40!black] (-0.6mm,4mm) node {$\Omega_2$};
\draw [red!60] (-3mm,2.5mm) node {$\omega_2$};
\draw [blue!40!black] (-4mm,0.5mm) node {$\Omega_3$};
\end{tikzpicture} 
\end{center}
\caption{The configuration of a tripartite entangling region for computing tripartite information, $\Omega_i$'s are the opening angles and $\omega_i$'s are the two smaller angular separations between them. The radial coordinate runs over $0\le \rho<\infty$.}
\label{fig:I3}
\end{figure}
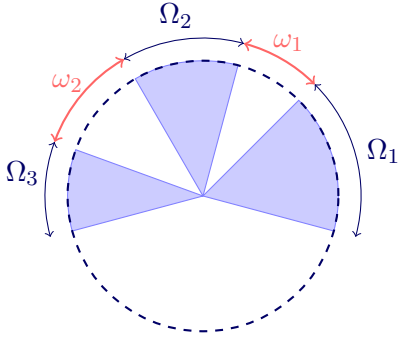

In this section we study the holographic tripartite information between kink entangling regions which are sectors of a single infinite circle. Tripartite information is defined as follows
\begin{align}\label{tripartite}
I^{[3]}(A_1,A_2,A_3)=S_{A_1}+S_{A_2}+S_{A_3}-S_{A_1\cup A_2}-S_{A_1\cup A_3}-S_{A_2\cup A_3}+S_{A_1\cup A_2\cup A_3},
\end{align}
where the $A_i$'s are entangling regions, which in our case are sectors of a single infinite circle with opening angles $\Omega_i,\;i=1, 2, 3$ and angular separation $\omega_j,\;j=1, 2$. Again we have excluded the maximum opening angle between three different choices to define $\omega_j$'s (see Fig.\ref{fig:I3}). In order to avoid the overlap between different regions we assume $\sum_{ij} (\Omega_i+\omega_j)< 2\pi$. Again as in the case of mutual information we reduce the parameter space from five independent parameters to two by considering the simplest case with $\Omega_i=\Omega, \omega_i=\omega$. 

\begin{figure}
\begin{center}
\includegraphics[scale=1]{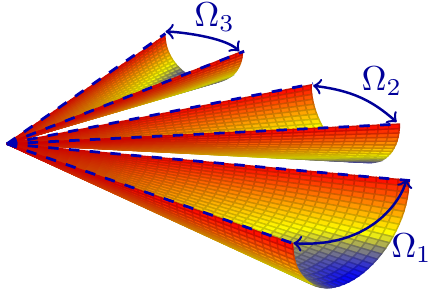}
\includegraphics[scale=1]{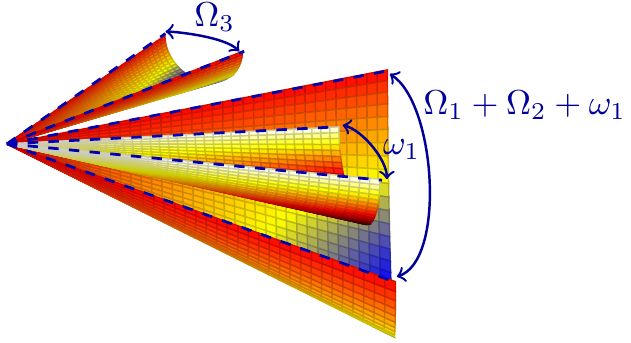}
\includegraphics[scale=1]{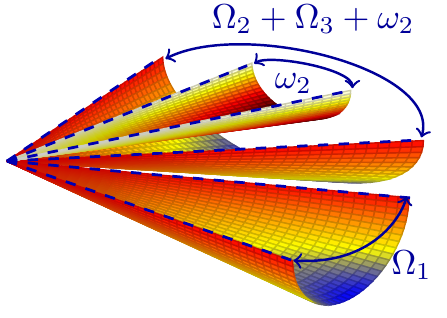}
\includegraphics[scale=1]{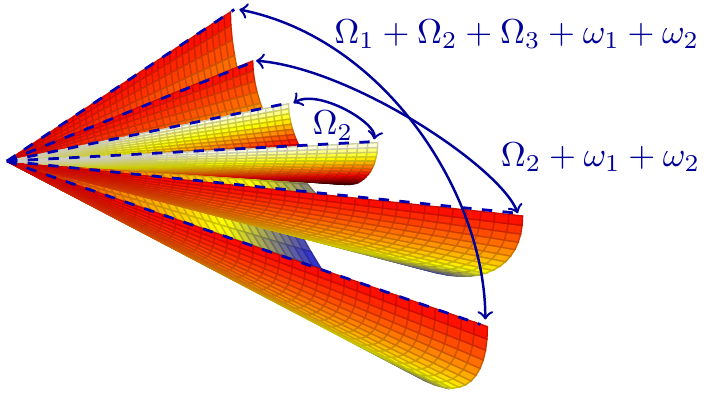}
\includegraphics[scale=1]{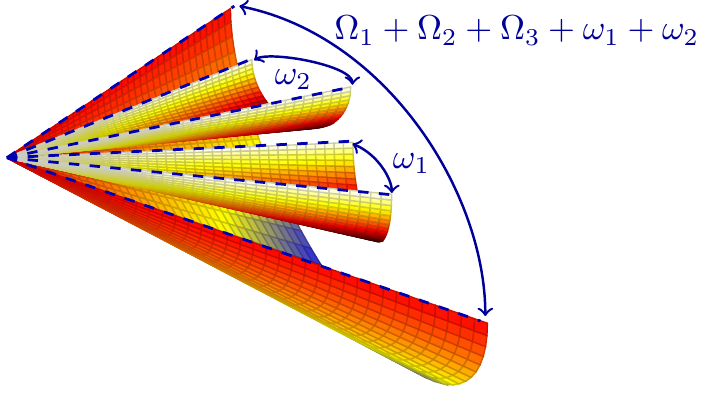}
\end{center}
\caption{Schematic representation of RT surfaces corresponding to $S_{\Omega_1\cup \Omega_2\cup \Omega_3}$ for computing the holographic tripartite information.}
\label{fig:I3config}
\end{figure}

\begin{figure}
\begin{center}
\includegraphics[scale=.55]{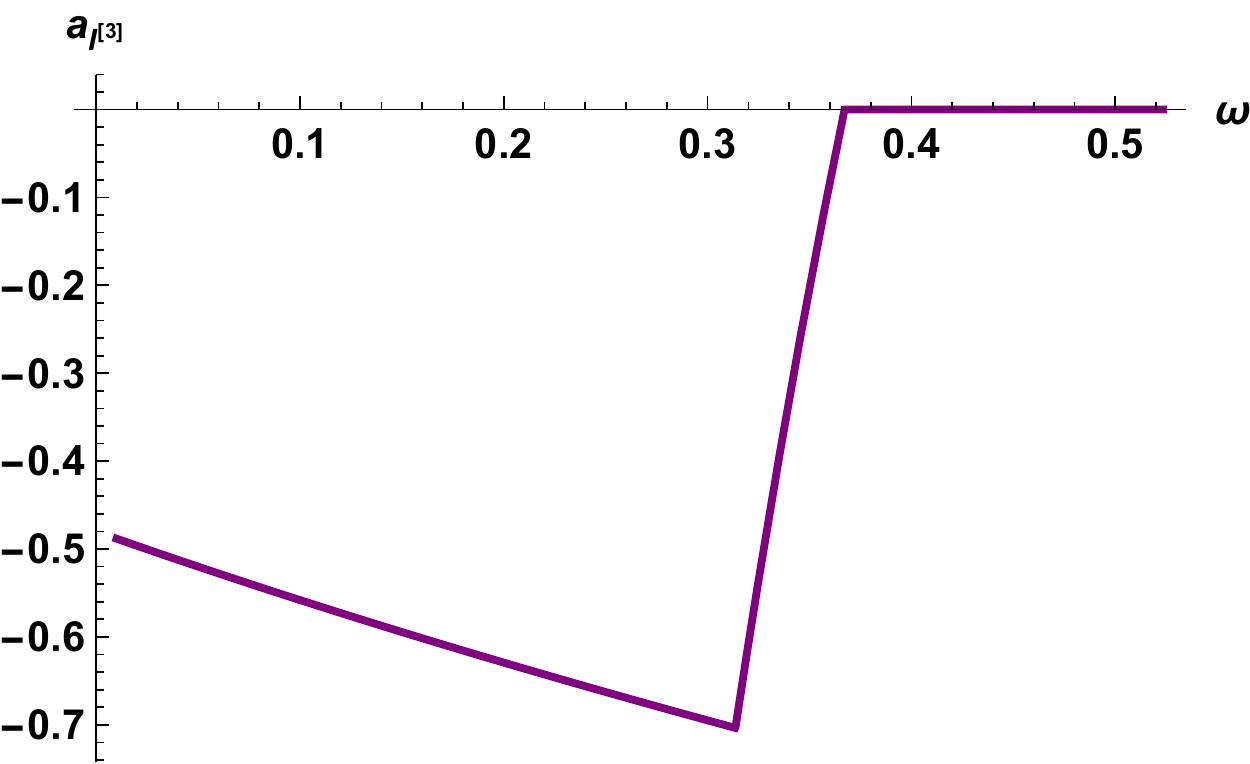}
\includegraphics[scale=.55]{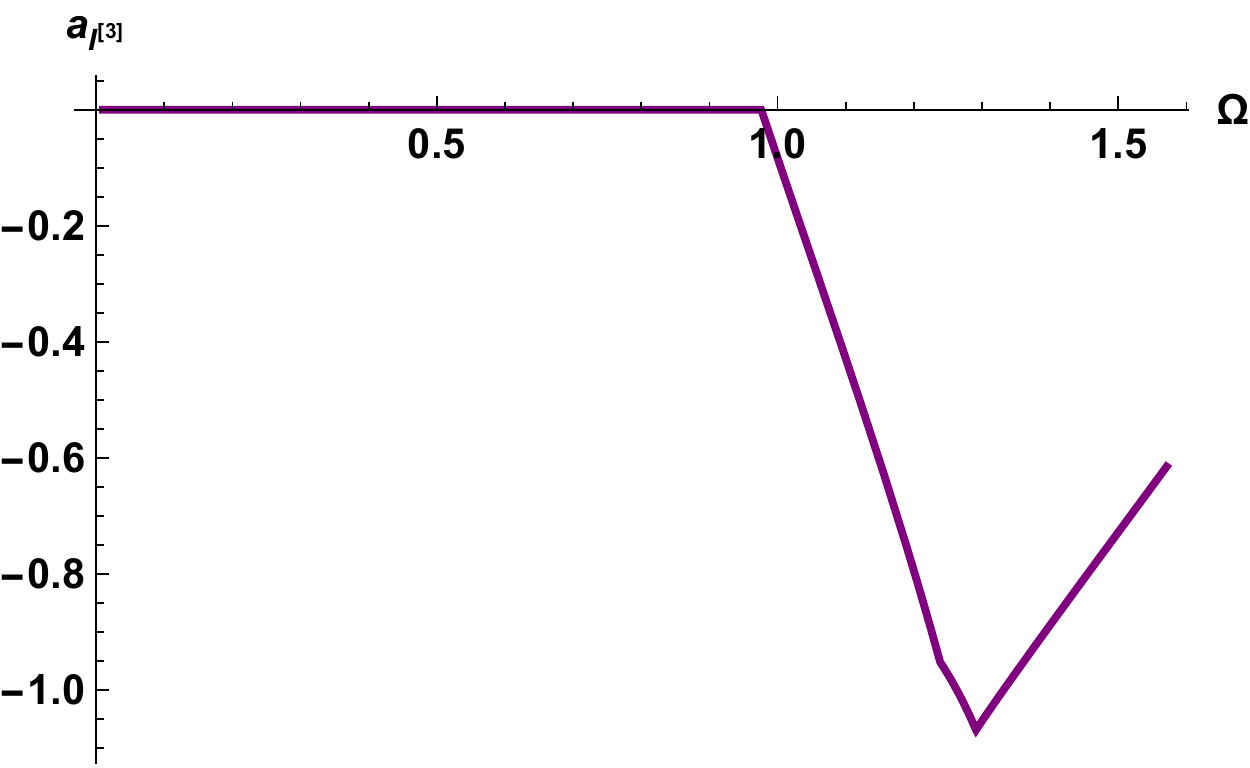}
\end{center}
\caption{Universal part of the holographic tripartite information for $\Omega=\frac{\pi}{6}$ (left) and $\omega=\frac{\pi}{4}$ (right) with $H=1$.}
\label{fig:tripartite}
\end{figure}

\begin{figure}
\begin{center}
\includegraphics[scale=.55]{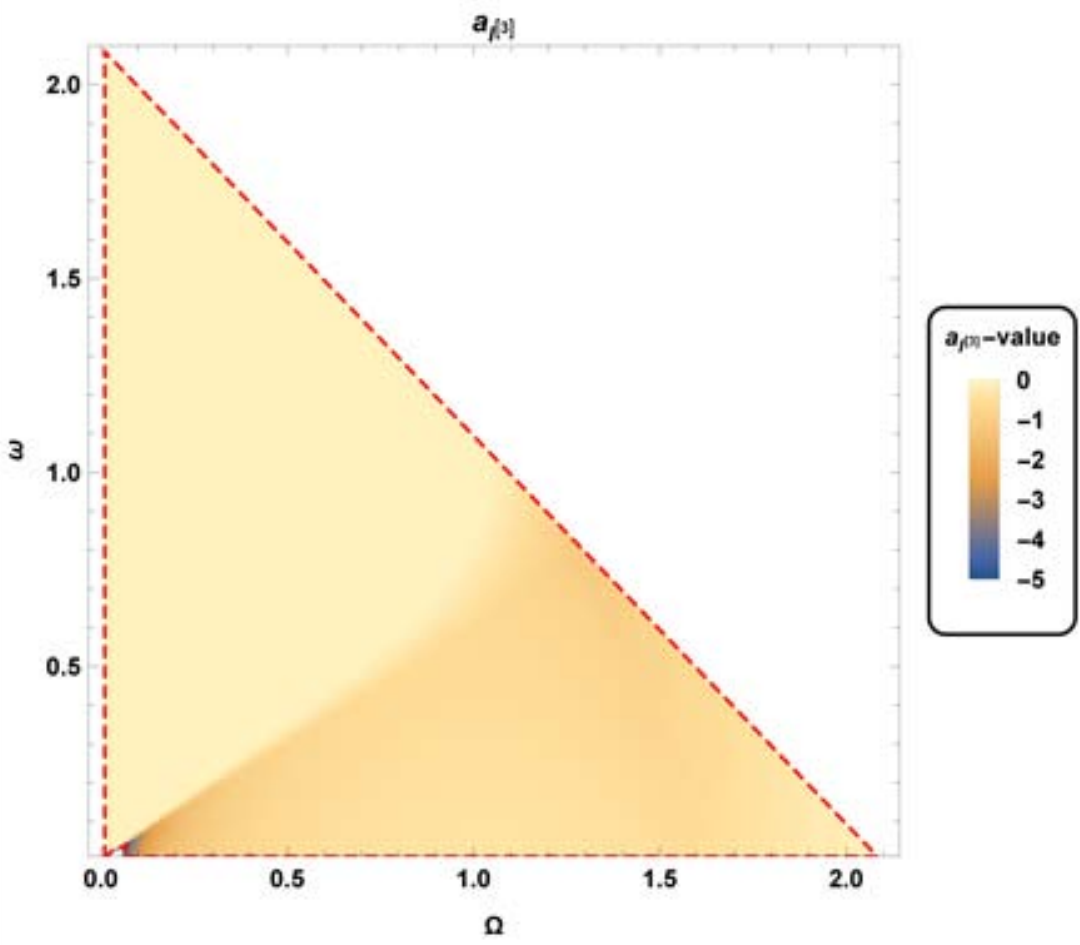}
\end{center}
\caption{Density plot of universal part of the holographic tripartite information for regions with equal opening angles. The dashed red boundary corresponds to $\Omega+\omega<\frac{2\pi}{3}$ constraint.}
\label{fig:densityplottripar}
\end{figure}

Here the main challenge to calculate holographic tripartite information is finding the minimal area for the union of subsystems. Similar to what we did in the previous subsection, here we can find the relevant configurations for the entanglement entropies appearing in Eq.\eqref{tripartite} as minimal surfaces of kink entangling regions with certain opening angles. These configurations are schematically plotted in Fig.\ref{fig:I3config}. In the extreme cases the minimal configurations are
\begin{align}
%S_{\Omega_2\cup \Omega_i}=\Bigg\{ \begin{array}{rcl}
%&S_{2\Omega+\omega}+S_{\omega}\equiv S^{(1)}_{\text{dis.}}&\,\,\,\,\,\omega \ll \Omega,\\
%&2S_{\Omega}\equiv S^{(2)}_{\text{dis.}}&\,\,\,\,\,\omega\gg \Omega,
%\end{array}\,\,\;\;\;\;\;i=1 \;\text{or}\; 3,
S_{\Omega_2\cup \Omega_i}=
\begin{cases}\label{union1}
S_{2\Omega+\omega}+S_{\omega}\left(\equiv S^{(1)}_{\text{dis.}}\right)  & ~~ \omega \ll \Omega\\
2S_{\Omega}\left(\equiv S^{(2)}_{\text{dis.}}\right) & ~~ \omega\gg \Omega
\end{cases}
\end{align}
for $i=1,3$ and 
\begin{align}\label{union2}
%S_{\Omega_1\cup \Omega_3}=\Bigg\{ \begin{array}{rcl}
%&2S_{\Omega}\equiv S^{(2)}_{\text{dis.}}&\,\,\,\,\,\omega\ll \Omega,\\
%&S_{3\Omega+2\omega}+S_{\Omega+2\omega}\equiv S^{(3)}_{\text{dis.}}&\,\,\,\,\,\omega\gg \Omega,
%\end{array}\,\,\;,
S_{\Omega_1\cup \Omega_3}=
\begin{cases}
2S_{\Omega}\left(\equiv S^{(2)}_{\text{dis.}}\right)  & ~~ \omega\ll \Omega\\
S_{3\Omega+2\omega}+S_{\Omega+2\omega}\left(\equiv S^{(3)}_{\text{dis.}}\right) & ~~ \omega\gg \Omega
\end{cases}
\end{align}
and for the case of union of three subsystems
\begin{align}\label{union3}
%S_{\Omega_1\cup \Omega_2\cup \Omega_3}=\Bigg\{ \begin{array}{rcl}
%&S_{3\Omega+2\omega}+2S_{\omega}\equiv S_{\text{con.}}&\,\,\,\,\,\omega\ll \Omega,\\
%&S_{2\Omega+\omega}+S_{\Omega}+S_{\omega}\equiv S^{(4)}_{\text{dis.}}&\,\,\,\,\,\\
%&3S_{\Omega}\equiv S^{(5)}_{\text{dis.}}&\,\,\,\,\,\omega\gg \Omega,\\
%&S_{3\Omega+2\omega}+S_{\Omega+2\omega}+S_{\Omega}\equiv S^{(6)}_{\text{dis.}}&\,\,\,\,\,
%\end{array}\,\,\;.
S_{\Omega_1\cup \Omega_2\cup \Omega_3}=
\begin{cases}
S_{3\Omega+2\omega}+2S_{\omega}\equiv S_{\text{con.}}  & ~~ \omega\ll \Omega\\
S_{2\Omega+\omega}+S_{\Omega}+S_{\omega}\equiv S^{(4)}_{\text{dis.}}  & ~~ \omega\gg \Omega\\
3S_{\Omega}\equiv S^{(5)}_{\text{dis.}}  & ~~ \omega\gg \Omega\\
S_{3\Omega+2\omega}+S_{\Omega+2\omega}+S_{\Omega}\equiv S^{(6)}_{\text{dis.}}  & ~~ \omega\gg \Omega
\end{cases}.
\end{align}
%
%
%
%\begin{align}\label{union2}
%S_{A_1\cup A_2}&=S_{A_2\cup A_3}=\min \{2S_\Omega, S_{2\Omega+\omega}\}\nonumber\\
%S_{A_2\cup A_3}&=\min \{2S_\Omega, S_{\Omega+2\omega}\},
%\end{align}
%and 
%\begin{align}\label{union3}
%&S({A_1\cup  A_2}\cup A_3)=\nonumber\\
%&\min \{3S(\Omega),S(2\Omega+\omega)+S(\Omega)+S(\omega),S(3\Omega+2\omega)+S(\Omega+2\omega)+S(\Omega), S(3\Omega+2\omega)+2S(\omega)\}.
%\end{align}
%In order to find the corner contributions to the holographic tripartite information we use the results of \cite{Bueno:2015xda} (section 2). In this case the HEE becomes
%\begin{align}\label{HEE}
%S_{EE}(\Omega)=\frac{L^2}{2G_N}\frac{H}{\epsilon}-a(\Omega)\log \frac{H}{\epsilon}-\left(\frac{\pi L^2}{4G_N h_*}+a(\Omega)\log h_*\right)+\mathcal{O}(\frac{\epsilon}{H}),
%\end{align}
%where $H$ and $\epsilon$ are infrared regulator and UV cut-off respectively. Also $h_*$ and $\Omega$ are turning point and opening angle respectively. The function $a(\Omega)$ is defined as follows
%\begin{align}\label{aomega}
%a(\Omega)=\frac{L^2}{2G_N}\int_0^\infty dy\left[1-\sqrt{\frac{1+h_*^2(1+y^2)}{2+h_*^2(1+y^2)}}\right].
%\end{align}
%Note that here we have two different kind of singularity, an area law and a new logarithmic divergence.
One can calculate the tripartite information by making use of Eq.\eqref{HEE} together with Eq.'s\eqref{tripartite}, \eqref{union1}, \eqref{union2} and \eqref {union3}. It is easy to check that the area law divergences cancel out in tripartite information but the logarithmic divergences does not cancel out and the resultant tripartite information is UV-divergent. This could be understood as a straightforward generalization of what was explained in the introduction section below Fig.\ref{fig:MIS} and where we referred to it. As an explicit example we have plotted the universal part of this quantity i.e. $a_{I^{[3]}}$, in Fig.\ref{fig:tripartite} (This universal part is defined similar to Eq.\eqref{aI} but using the definition of tripartite information). Fig.\ref{fig:densityplottripar} also shows the density plot of this quantity. According to these plots the holographic tripartite information in this singular set-up is negative as expected. 
Using the small opening angle expansion $\{\Omega,\omega\}\rightarrow 0$ one can also study the behavior of tripartite information semi-analytically. In this case one finds
\begin{align}
I^{[3]}(\Omega, \omega)=\begin{cases}
-\frac{\kappa}{3\Omega}\log\frac{H}{\epsilon}& ~~ \omega\ll \Omega\\
0  & ~~ \omega\gg \Omega\\
\end{cases},
\end{align}
which the final result still depends on the inverse UV cut-off $\epsilon$. In order to derive this result we use the similar analysis as in deriving Eq.\eqref{Ismall}.
% See for example Fig. \ref{fig:3partie-smallcone}. 
This behavior is in contrast with the well-known feature of tripartite information in the literature, which is thought to be finite even when the regions share boundaries \cite{Hayden:2011ag}. See appendix \ref{sec:app1} for some details about the finiteness of tripartite information for smooth entangling regions in various dimensions. 

%\begin{figure}
%\begin{center}
%\includegraphics[scale=.8]{I31001000}
%\end{center}
%\caption{The tripartite information for $\omega=\frac{\pi}{2}-\Omega$ as a function of $\Omega$ for three different values of $H/\epsilon=10^3, 10^6, 10^{12}$.}
%\label{fig:EEMI}
%\end{figure}
%\begin{figure}
%\begin{center}
%\includegraphics[scale=.8]{3partie-smallcone}
%\end{center}
%\caption{The tripartite information in the small opening angle limit for $\omega=\frac{\pi}{12}$ as a function of $\Omega$ for three different values of $H/\epsilon=10^3, 10^6,10^{12}$.}
%\label{fig:3partie-smallcone}
%\end{figure}
\subsection{Holographic $n$-partite Information}
In this section we generalise our previous study to the case of holographic $n$-partite information. This quantity is a simple generalization of mutual and tripartite information to systems consisting of $n$ disjoint subsystems. The definition of $n$-partite information is as follows \cite{Hayden:2011ag}
\begin{align}\label{eq:npartite}
I^{[n]}(A_{\{i\}})=\sum_{i=1}^nS_{A_i}-\sum_{i<j}^n S_{A_i\cup A_j}+\sum_{i<j<k}^n S_{A_i\cup A_j\cup A_k}
-\cdots\cdots -(-1)^n S_{A_1\cup A_2\cup\cdots\cup A_n},
\end{align}
where for $n=2$ and $n=3$ it reduces to the definition of mutual and tripartite information.\footnote{Actually the generalisation to generic $n$ is not unique, but we choose this definition to reproduce the tripartite information for $n=3$. Other definition e.g., multipartite-information does not satisfy this constraint\cite{Horodecki:2009zz}.} This quantity is finite and in a general quantum system it can be either negative, positive or zero. Holographic computations show that in certain limits it has a definite sign, positive (negative) for even (odd) $n$\cite{Alishahiha:2014jxa}.

\begin{figure}
\begin{center}
\includegraphics[scale=.55]{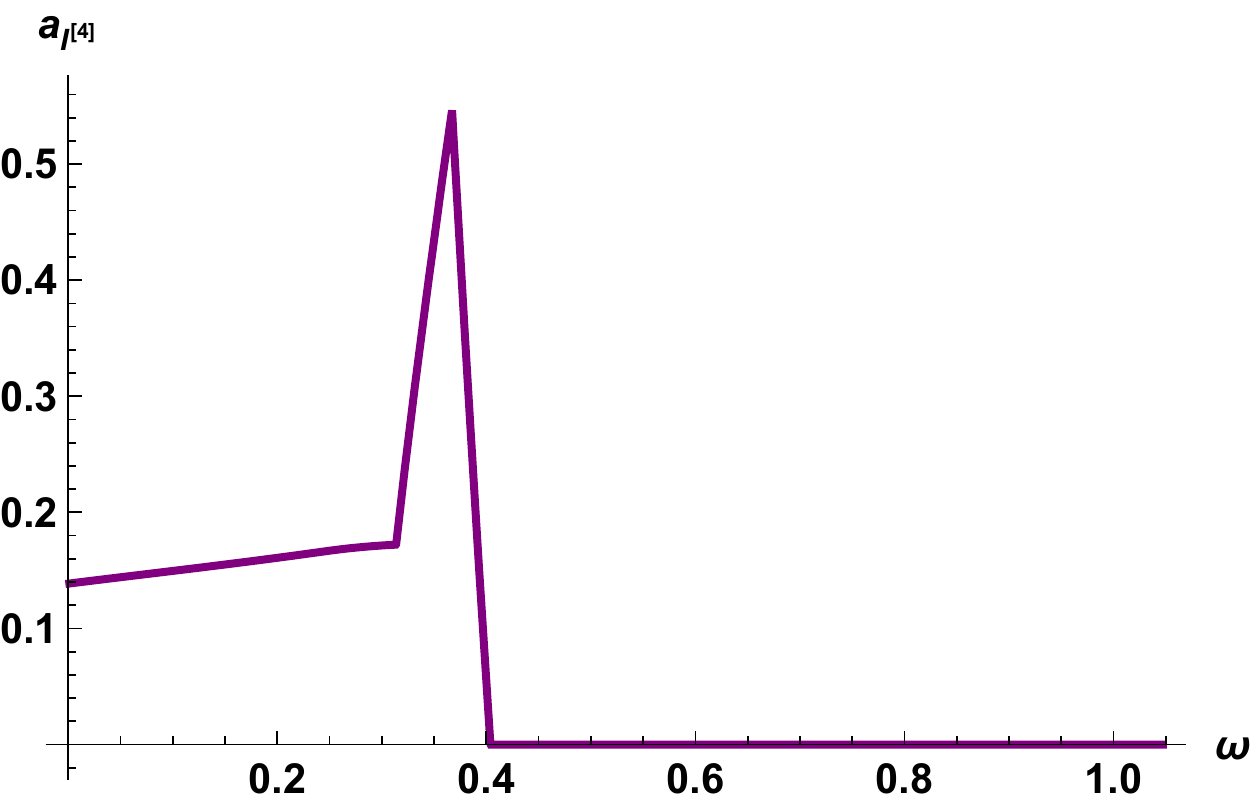}
\includegraphics[scale=.55]{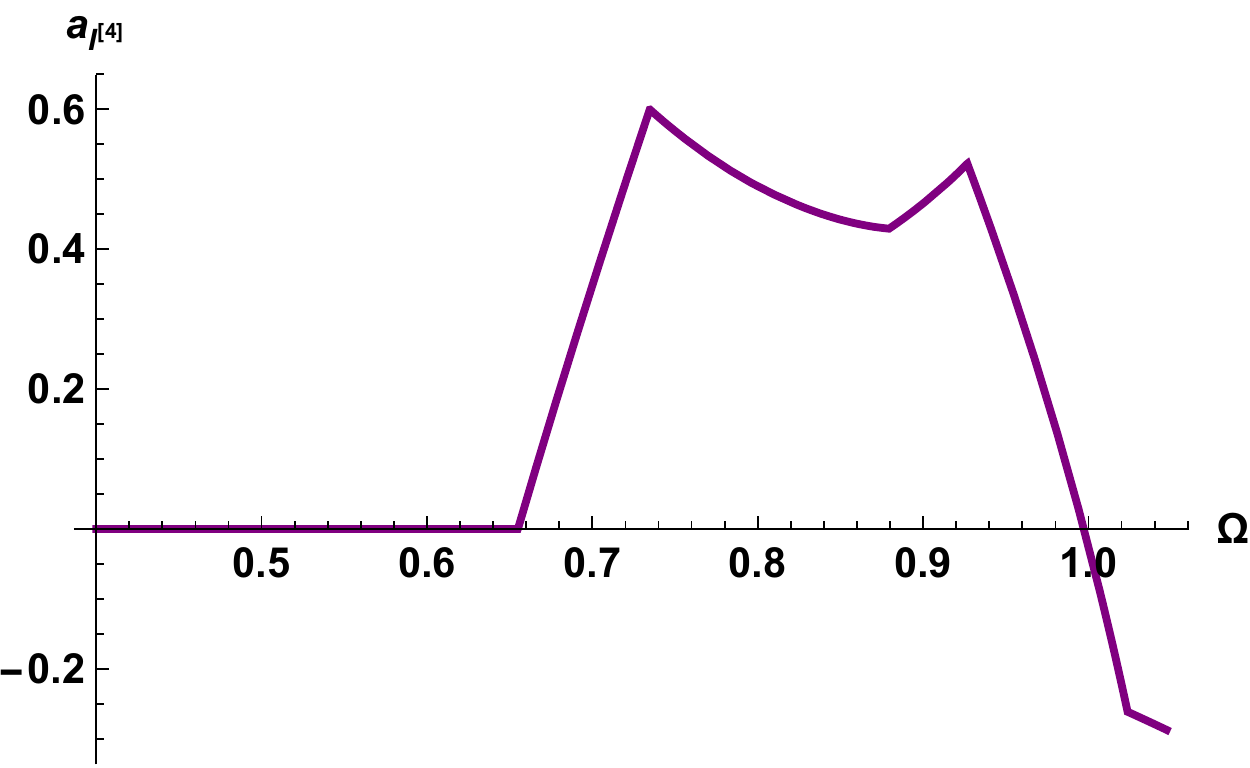}
\end{center}
\caption{Universal part of the holographic 4-partite information for $\Omega=\frac{\pi}{6}$ (left) and $\omega=\frac{\pi}{6}$ (right) with $H=1$.}
\label{fig:4partie}
\end{figure}

In order to simplify the computations we only consider the case where all the opening angles and separations between them are equal, i.e., $\Omega_1=\Omega_2=\cdots=\Omega_n\equiv\Omega$ and $\omega_1=\omega_2=\cdots=\omega_{n-1}\equiv\omega$. For such a choice similar to the case of mutual information and tripartite information we have a geometric constraint $\Omega+\omega<2\pi/n$. Actually the analysis for finding the $n$-partite information in this singular set-up is very similar to the case of strip entangling regions \cite{Alishahiha:2014jxa} so we just demonstrate the final results (similar analysis for computing HEE for multiple strips has been done in \cite{Ben-Ami:2014gsa}). The simplest example is $n=4$ which corresponds to 4-partite information. Fig.\ref{fig:4partie} shows the behavior of the universal part of holographic 4-partite information as a function of $\Omega$ and $\omega$. Also in the small opening angle limit $\{\Omega,\omega\}\rightarrow 0$ one finds
\begin{align}
I^{[4]}(\Omega, \omega)=\begin{cases}
\frac{\kappa}{12\Omega}\log\frac{H}{\epsilon}& ~~ \omega\ll \Omega\\
0  & ~~ \omega\gg \Omega\\
\end{cases}.
\end{align}
One can also find the small angle expression for the holographic $n$-partite information for generic $n$ as follows \footnote{Note that similar to previous cases, in order to avoid the overlap between different regions we assume $\sum_{ij}\Omega_i+\omega_j< 2\pi$ where $i=1,\cdots, n$ and $j=1,\cdots, n-1$.}
\begin{align}
I^{[n]}(\Omega \sim 0, \omega\sim 0)=\begin{cases}
(-1)^n\frac{2\kappa}{n(n-1)(n-2)\Omega}\log\frac{H}{\epsilon}& ~~ \omega\ll \Omega\\
0  & ~~ \omega\gg \Omega\\
\end{cases}.
\end{align}
The above result shows that in this specific construction, considering small separation angles, the holographic $n$-partite information has a definite sign, i.e., it is positive (negative) for even (odd) $n$. Note that using the conformal map which is given by Eq.\eqref{cmap} and the result of $n$-partite information for a set of strips with equal width $\ell$ and separation $x$, considering the $\ell\gg x$ limit (see \cite{Alishahiha:2014jxa}) one can reproduce the above result. 

\section{Holographic Entanglement Measures in Higher Dimensions}\label{sec:HD}

In this section we generalise our studies to higher dimensional cases in a specific direction where the entangling region is a crease, i.e., $k\times R^m$ (see the right panel of Fig.\ref{fig:kc}). To do so we first review the computation of HEE for crease entangling regions and continue with computing the holographic mutual and tripartite information for this generalization of the specific configurations we considered in the previous section. The HEE of such entangling regions has been studied in \cite{Myers:2012vs}.

The dual geometry we consider here is an AdS$_{d+1}$ space-time with the following metric in cylindrical coordinates which is the specific case of $n=0$ and $m=d-3$ of Eq.\eqref{metric1}
\begin{align}
ds^2=\frac{L^2}{z^2}\left(dz^2-dt^2+d\rho^2+\rho^2 d\theta^2+\sum_{i=1}^{d-3}\;dx_i^2\right).
\end{align}
We are interested in the following entangling region
\begin{equation}
t=\mathrm{const.}\;\;\;,\;\;\;0<\rho< H\;\;\;,\;\;\;-\frac{\Omega}{2}\leq\theta \leq\frac{\Omega}{2},\;\;\;0<x_i<\tilde{H},
\end{equation}
where both $H$ and $\tilde{H}$ are IR regulators. Due to the symmetries we assume $z=z(\rho, \theta)$, thus the HEE functional becomes
\begin{align}\label{higherdim}
S=\frac{L^{d-1}\tilde{H}^{d-3}}{4G_N}\int d\rho d\theta \frac{\sqrt{\dot{z}^2+\rho^2+\rho^2{z'}^2}}{z^{d-1}},
\end{align}
%\begin{align}\label{higherdim}
%S=\frac{L^{d-1}\tilde{H}^{d-3}}{2G_N}\int_{\frac{\epsilon}{h_*}}^H \frac{d\rho}{\rho^{d-2}}\int_{0}^{\frac{\Omega}{2}-\delta}d\theta\frac{\sqrt{1+h^2+{h'}^2}}{h^{d-1}},
%\end{align}
where $z'=\partial_\rho z$ and $\dot{z}=\partial_\theta z$. Similar to the three dimensional case, using the scaling symmetry, here we consider $z(\rho, \theta)=\rho\;h(\theta)$ and find a conserved quantity as follows \cite{Myers:2012vs}
\begin{align}\label{Hd}
\mathcal{H}_d\equiv \frac{(1+h^2)^{\frac{d-1}{2}}}{h^{d-1}\sqrt{1+h^2+{h'}^2}}=\frac{(1+h_*^2)^{\frac{d-2}{2}}}{h_*^{d-1}}.
\end{align}
Using this equation the opening angle $\Omega$ is related to the turning point as
\begin{align}
\Omega=\int_0^{h_*}dh \frac{2\mathcal{H}_dh^{d-1}}{\sqrt{1+h^2}\sqrt{(1+h^2)^{d-2}-\mathcal{H}_d^2h^{2(d-1)}}}.
\end{align}
\begin{figure}
\begin{center}
\includegraphics[scale=.65]{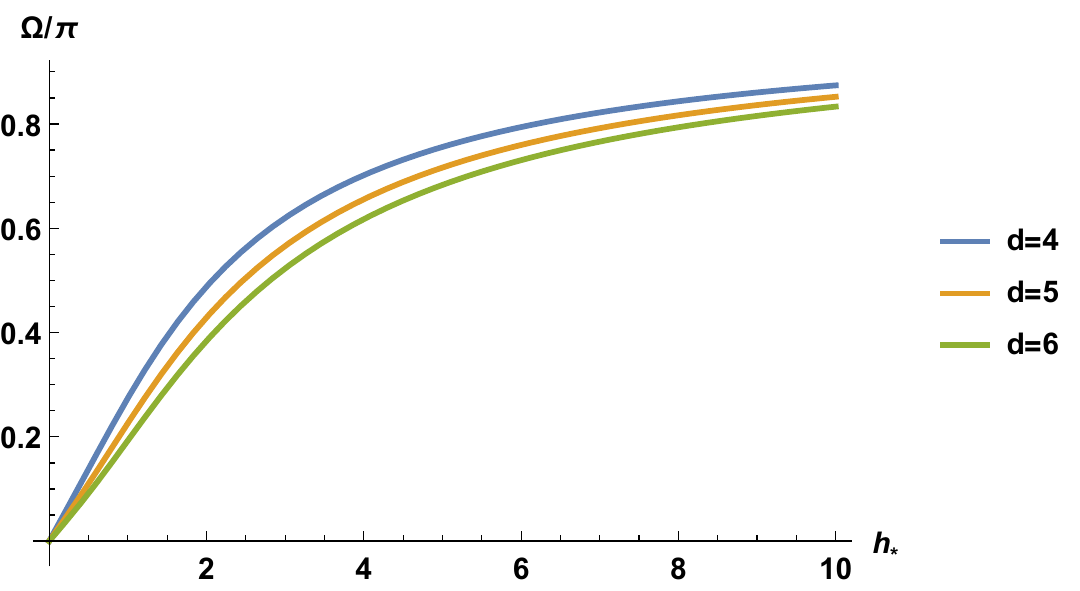}
\includegraphics[scale=.65]{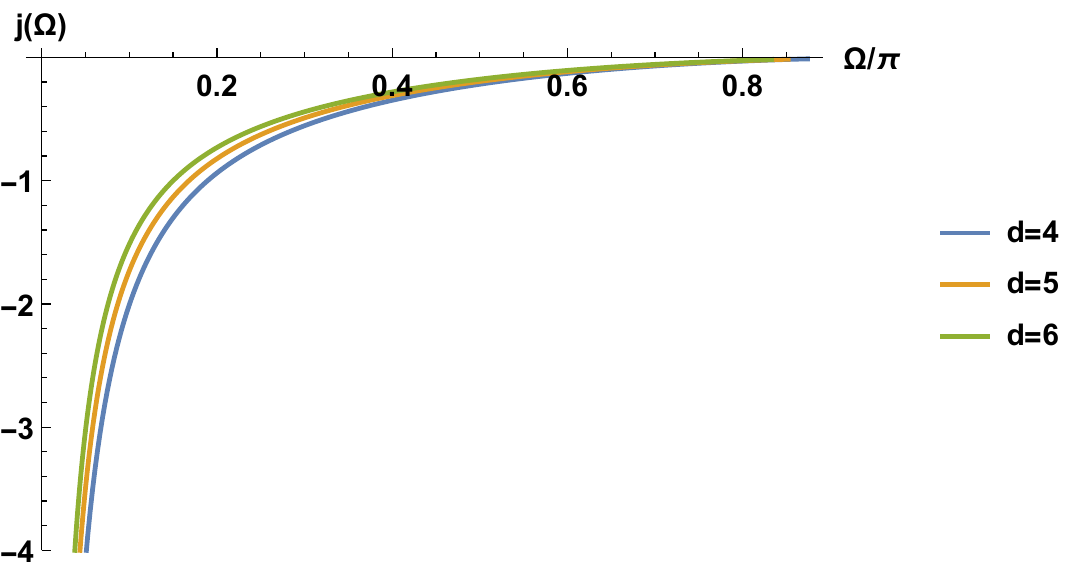}
\end{center}
\caption{\textit{Left}: $\Omega/\pi$ as a function of the turning point $h_*$ in different dimensions. \textit{Right}: $j$ as a function of the opening angle $\Omega$ in different dimensions.}
\label{fig:highd}
\end{figure}
The HEE could be found as
\begin{align}\label{EEhighd}
S=\frac{L^{d-1}\tilde{H}^{d-3}}{2G_N}\left[\frac{H}{(d-2)\epsilon^{d-2}}+\frac{j(\Omega)}{(d-3)\epsilon^{d-3}}\right]+\mathcal{O}(\epsilon),
\end{align}
where
\begin{align}
j(\Omega)=\int_{h_*}^0 dh\;h^{d-3}J(h)-\frac{1}{h_*}\;\;\;\;,\;\;\;\;J(h)=\frac{\sqrt{1+h^2+{h'}^2}}{h'h^{d-1}}+\frac{1}{h^{d-1}},
\end{align}
noting that $h'$ is determined in terms of $h$ and $h_*$ via Eq.\eqref{Hd}. Eq.\eqref{EEhighd} shows that the logarithmic divergent term in the previous section was a reminiscent of three dimensional field theory and in higher dimensions a new power law divergent term appears. It is important to note that this new power law divergence, i.e., $\epsilon^{-(d-3)}$ does not appear if we consider smooth entangling regions. It is known from reference \cite{Myers:2012vs} that this behavior is due to adding a flat locus, i.e., $R^m$ to the kink $k$.
%If the and by turning the curvature of some even dimensional locus again the logarithmic divergence terms appear.
Fig.\ref{fig:highd} demonstrates the behavior of $\Omega(h_*)$ and $j(\Omega)$ in higher dimensions. From this plot one can deduce that $j(\Omega)$ vanishes in the smooth limit as expected.

Having the expression for the HEE, in principle we can compute the mutual information Eq.\eqref{mutual} between higher dimensional creases holographically. The subtlety of finding the minimal surface for union of the creases arises again in this case. The difficulty is similar to what happens while dealing with multi-strip entangling regions in higher dimensions. Remember that for a line segment entangling region in a two dimensional theory, the corresponding RT surface is a semi-circle ($s^1$) in an AdS$_3$ geometry. While we are dealing with an infinite strip in AdS$_{d+1}$, the corresponding RT surface is $\tilde{s}^1\times R^{d-2}$ where $\tilde{s}$ refers to a modified semi-circle \cite{Ryu:2006ef}. Using this simple generalization, the authors of \cite{Fischler:2012uv} have found the HMI for strip entangling regions in higher dimensions. However it is not possible to imagine the embedding of the corresponding RT surface for $S_{A_1\cup A_2}$ in the bulk for $d>3$. In the case of our study, where we generalize a kink $k$ in $d=3$ to a crease $k\times R^{d-3}$ in higher dimensions, again there is enough symmetry leading to a conserved quantity (see Eq.\eqref{Hd}). So we expect that the corresponding RT surfaces for computing the HEE for union of crease regions are just simple generalisation of three dimensional case. Note that for disk or spherical entangling regions this procedure does not work any more \cite{Fonda:2014cca}. This makes us to conclude that if we had considered a cone (instead of a crease) as a generalization of a kink in three dimensions, it would be impossible to construct the corresponding RT surfaces analytically.

\begin{figure}
\begin{center}
\includegraphics[scale=.7]{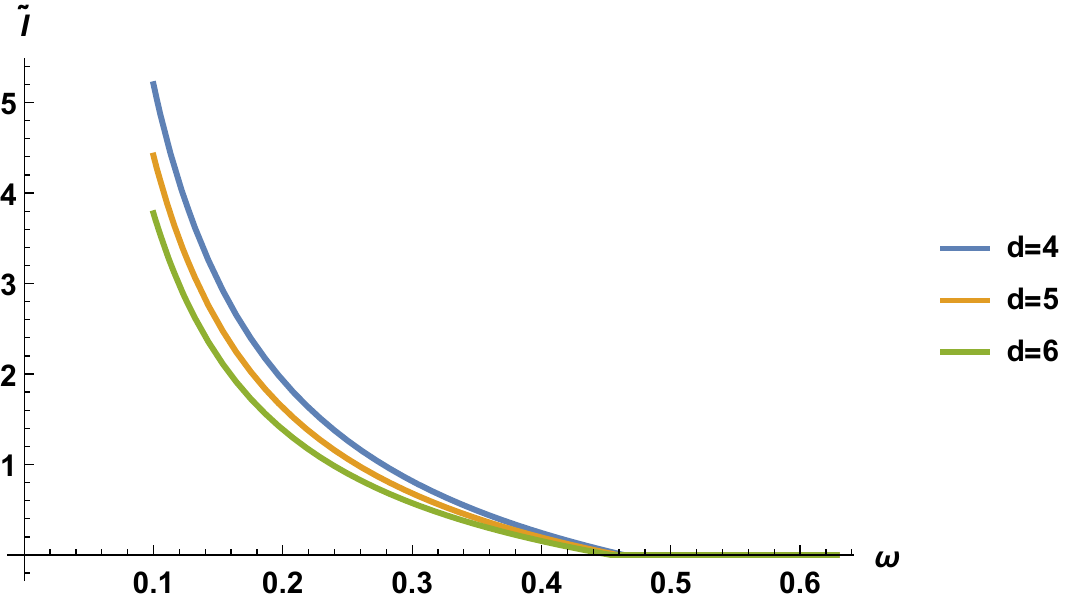}
\includegraphics[scale=.7]{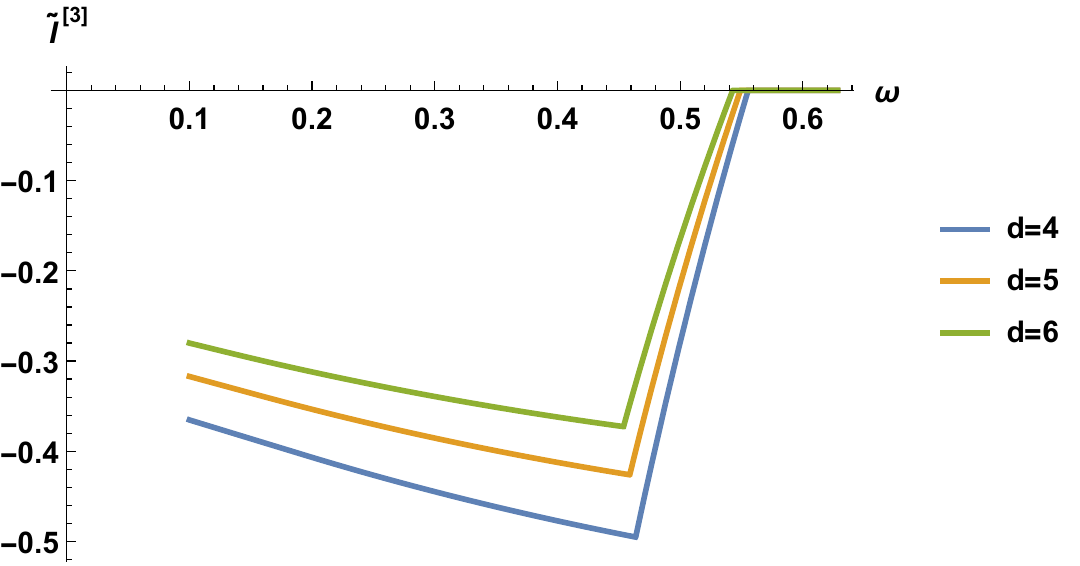}
\end{center}
\caption{Normalized holographic mutual (left) and tripartite (right) information in various spatial dimensions for $\Omega=\frac{\pi}{4}$.}
\label{fig:HMIhighd}
\end{figure}

Now using the above argument one can find the HMI in our set-up when the opening angles and separation angle are given by $\Omega_1,\;\Omega_2$ and $\omega$ respectively. Doing so the area law divergences trivially cancel out in the expression of HMI and we are left with
\begin{align}
I(\Omega_1, \Omega_2)=\frac{L^{d-1}\tilde{H}^{d-3}}{2G_N}\left[j(\Omega_1)+j(\Omega_2)-\min \left\{j(\Omega_1)+j(\Omega_2),j(\Omega_1+\Omega_2+\omega)+j(\omega)\right\}\right]\frac{1}{(d-3)\epsilon^{d-3}}.
\end{align}
In order to simplify the computations we only consider the equal opening angle case, where the resultant HMI is plotted in Fig.\ref{fig:HMIhighd} for various spatial dimensions. In this figure we have plotted a normalized holographic mutual information defined as $\tilde{I}=(d-3)\epsilon^{d-3}I$. This figure shows that the transition always exists, however it depends on UV cut-off similar to the three dimensional case. Similar analysis can be done in the case of holographic tripartite information, see Fig.\ref{fig:HMIhighd}. Also note that in order to plot this figure we have set $\frac{L^{d-1}\tilde{H}^{d-3}}{2G_N}=1$.

%\begin{figure}\label{fig:I3}
%\begin{center}
%\begin{tikzpicture}[scale=6]
%  \fill[fill=blue!20!white,rotate=-45, draw=blue!50!white]
%    (0,0) -- (3mm,0mm) arc (0:30:3mm) -- (0,0);
%  \fill[fill=blue!20!white,rotate=75, draw=blue!50!white]
%    (0,0) -- (3mm,0mm) arc (0:30:3mm) -- (0,0);
%  \fill[fill=blue!20!white,rotate=-165, draw=blue!50!white]
%    (0,0) -- (3mm,0mm) arc (0:30:3mm) -- (0,0);
%        \draw[thick,blue!40!black] (0cm,0cm) circle(3mm);    
%%%%%%%%%%%%%%%%%%%
%  \draw[blue!40!black,<->, rotate=-15]
%    (3.5mm,0mm) arc (0:30:3.5mm);
%  \draw[red!60,thick,<->, rotate=15]
%    (3.5mm,0mm) arc (0:60:3.5mm);
%  \draw[blue!40!black,<->, rotate=75]
%    (3.5mm,0mm) arc (0:30:3.5mm);
%  \draw[red!60,thick,<->, rotate=105]
%    (3.5mm,0mm) arc (0:60:3.5mm);
%  \draw[blue!40!black,<->, rotate=165]
%    (3.5mm,0mm) arc (0:30:3.5mm);
%\draw [blue!40!black] (4.2mm,0) node {$\Omega$};
%\draw [red!60] (3.6mm,2.6mm) node {$\frac{\pi}{2}-\Omega$};
%\draw [blue!40!black] (0,4.2mm) node {$\Omega$};
%\draw [red!60] (-3.6mm,2.6mm) node {$\frac{\pi}{2}-\Omega$};
%\draw [blue!40!black] (-4.2mm,0) node {$\Omega$};
%\end{tikzpicture} 
%\end{center}
%\caption{Three parts of the entangling region have all equal arc widths, $\Omega$ and the radial coordinate is $0\le \rho<\infty$. }
%\end{figure}
\section{Mutual Information Between Sharp Concentric Circles}\label{sec:SCC}

In the previous sections we studied some entanglement measures including mutual information for configurations which the entangling regions had a common point which was the same as their singular point. We argued in the introduction section (see the caption of Fig.\ref{fig:MIS} and where we referred to it) that for such configurations we do not expect mutual information and even tripartite information to be UV-finite quantities. What if we study e.g. mutual information between singular surfaces which do not share a boundary? In such a case we expect the result to be again a UV-finite quantity.
   
In this section we are going to explore another explicit example for mutual information between singular surfaces in holographic three dimensional conformal field theories. We aim to provide evidence for UV cut-off independence of the mutual information in three dimensions between two singular regions without a common boundary.
%The singular regions we consider here do not have a contact point in contrast with what was studied in the previous section. 

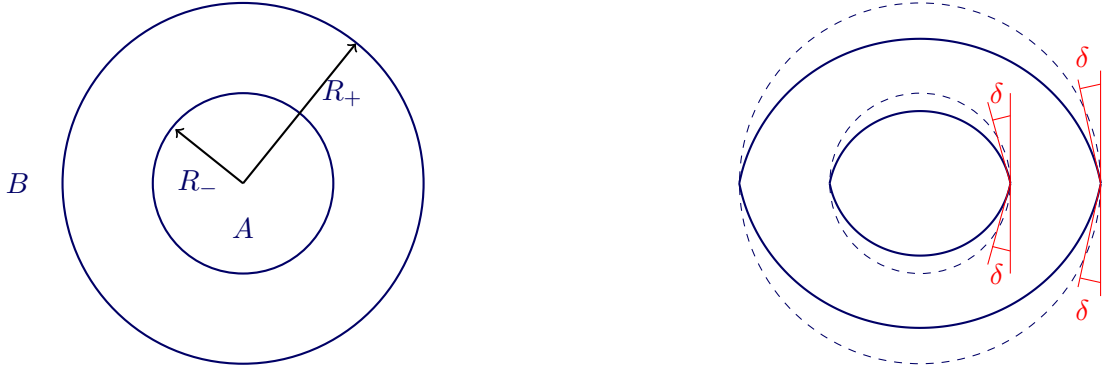
\begin{figure}
\begin{center}
\begin{tikzpicture}[scale=6]
\draw[blue!40!black,dashed] (0cm,0cm) circle(2mm);    
\draw[blue!40!black,dashed] (0cm,0cm) circle(4mm);    
%%%%%%%%%%%%%%%%%%%%%%%%%
%%%%%%%%%%%%%%%%%%%%%%%%%
\draw[thick,blue!40!black] (-1.5cm,0cm) circle(2mm);    
\draw[thick,blue!40!black] (-1.5cm,0cm) circle(4mm);    
\draw[->,thick] (-15mm,0)--(-16.5mm,1.2mm);
\draw[->,thick] (-15mm,0)--(-12.5mm,3.1mm);
\draw [blue!40!black] (-1.5cm,-1mm) node {$A$};
\draw [blue!40!black] (-20mm,0mm) node {$B$};
\draw [blue!40!black] (-16mm,0mm) node {$R_-$};
\draw [blue!40!black] (-12.8mm,2mm) node {$R_+$};
%%%%%%%%%%%%%%%%%%%%%%%%%
\draw[color=blue!40!black,thick,domain=0:3.14,samples=50,smooth] plot (canvas polar
cs:angle=\x r,radius= {5.7*1-5.7*.2*sin(\x r)});
\draw[color=blue!40!black,thick,domain=3.14:6.28,samples=50,smooth] plot (canvas polar
cs:angle=\x r,radius= {5.7*1+5.7*.2*sin(\x r)});
\draw[color=blue!40!black,thick,domain=0:3.14,samples=50,smooth] plot (canvas polar
cs:angle=\x r,radius= {11.4*1-11.4*.2*sin(\x r)});
\draw[color=blue!40!black,thick,domain=3.14:6.28,samples=50,smooth] plot (canvas polar
cs:angle=\x r,radius= {11.4*1+11.4*.2*sin(\x r)});
\draw [red](.2cm,-.2cm)--(.2cm,.2cm);
\draw [red](.2cm,0)--(.15cm,.18cm);
\draw [red](.16cm,.14cm)--(.2cm,.15cm);
\draw [red] (.17cm,.2cm) node {$\delta$};
\draw [red](.2cm,0)--(.15cm,-.18cm);
\draw [red](.16cm,-.14cm)--(.2cm,-.15cm);
\draw [red] (.17cm,-.2cm) node {$\delta$};
\draw [red](.4cm,-.25cm)--(.4cm,.25cm);
\draw [red](.4cm,0)--(.35cm,.23cm);
\draw [red](.4cm,.22cm)--(.355cm,.21cm);
\draw [red] (.36cm,.28cm) node {$\delta$};
\draw [red](.4cm,0)--(.35cm,-.23cm);
\draw [red](.4cm,-.22cm)--(.355cm,-.21cm);
\draw [red] (.36cm,-.28cm) node {$\delta$};
\end{tikzpicture} 
\end{center}
\caption{Left: Regions $A$ and $B$ which are separated by an annular region. Right: Deformation of the annular region with two singular points at the poles.}
\label{fig:annular}
\end{figure}

Consider two entangling regions $A$ and $B$ separated by an annular region again in the vacuum state of a three dimensional CFT. Region $A$ is a disk with radius $R_-$ and $B$ is the outer region (complement) of a larger disk with radius $R_+$ (see left panel of Fig.\ref{fig:annular}). Mutual information for such a configuration has been studied in details for certain field theories and also for holographic field theories in \cite{Nakaguchi:2014pha}. It would be interesting to consider a slight deformation on the annular region in-between these regions such that it brings in two corners for each entangling region with the same opening angle (see the right panel of Fig.\ref{fig:annular}). The question is whether the mutual information between $A$ and $B$ is still cut-off dependent as a result of the corners appeared perturbatively or not?

We again start with an AdS$_4$ space-time in the Poincare patch as the holographic dual of the three dimensional CFT
\be
ds^2=\frac{1}{z^2}\left(-dt^2+dz^2+d\rho^2+\rho^2d\phi^2\right)
\ee
and consider the boundary of the entangling regions defined as
$$t=\mathrm{const.}\;\;\;,\;\;\;\rho=R_{\pm}\;\;\;,\;\;\;0\le\phi<2\pi.$$
Following the method developed to consider generic perturbations on a spherical entangling region in \cite{Mezei:2014zla, Allais:2014ata} and also \cite{Bueno:2015lza} where the entanglement entropy of a particular choice of such deformations is analysed,  we consider the following perturbations on the boundaries of the entangling region
\be
\frac{\rho_{\pm}(\phi)}{R_{\pm}}=
\begin{cases}
1-\delta\,\sin\left(n\,\phi\right)&~~ 0\le\phi<\pi\\
1+\delta\,\sin\left(n\,\phi\right)&~~ \pi\le\phi<2\pi
\end{cases}
\ee
where $n$ is a positive integer number controlling the geometry of the perturbed entangling region and $\delta$ is a small parameter controlling the perturbation. Such a generic perturbation makes it possible to study various types of entangling regions but since we are interested in the specific deformation illustrated in the right panel of Fig.\ref{fig:annular}, from now on we restrict our analysis to the case $n=1$.
%See Fig.\ref{fig:ExAnn} for some examples.
%\begin{figure}
%\begin{center}
%\includegraphics[scale=.7]{annularEx1}
%\hspace{1.5cm}
%\includegraphics[scale=.7]{annularEx2}
%\end{center}
%\caption{The region between blue boundaries is purturbed to the region between orange boundaries with (Left:) $a_<=a_>=b=1$ and (Right:) $a_<=a_>=1$ and $b=2$. We have chosen $\delta=0.2$.}
%\label{fig:ExAnn}
%\end{figure}

In order to study the entanglement entropy holographically for such deformed regions and thus compute the mutual information between them, we consider a minimal surface in the bulk parametrized as $\rho(z,\phi)$. The induced metric on such an extended surface in the bulk geometry is
\be
\mathcal{A}=\int dz d\phi\frac{1}{z^2}\sqrt{\rho^2\left(1+{\rho'}^2\right)+\dot{\rho}^2}
\ee
where prime and dot indicate differentiation with respect to $z$ and $\phi$ respectively. The above functional could be minimized up to $\mathcal{O}\left(\delta^2\right)$ via 
\be
\rho(z,\phi)=\rho_0(z)+\delta\rho_1(z)\sin\phi+\delta^2\left[\rho_{20}(z)+\rho_{22}(z)\cos\left(2\phi\right)\right],
\ee
where $\rho_0(z)$, $\rho_1(z)$ and $\rho_{20}(z)$ must satisfy the following equations
\begin{align}\label{eq:deq}
\begin{split}
0&=z\left(1+{\rho'_0}^2\right)^2+\rho_0\left(2\rho'_0+2{\rho'_0}^3-z\rho_0^{''}\right)\\
0&=2\rho_0\left(\rho_0+z\rho'_0+3\rho_0{\rho'_0}^2\right)\rho'_1-z\rho_0^2\rho_1^{''}\\
0&={\rho_0'}^2\rho_1^2\left(z+\rho_0\rho'_0\right)-z\rho_0\rho_1\rho'_0\rho'_1-\frac{1}{2}\rho_0^2\left(z+6\rho_0\rho'_0\right){\rho_1'}^2+z\rho_0\left(1+{\rho'_0}^2\right)\rho_{20}\\
&\hspace{4mm}-2\rho_0^2\left(\rho_0+z\rho_0'+3\rho_0{\rho_0'}^2\right)\rho_{20}'+z\rho_0^3\rho_{20}^{''}.
\end{split}
\end{align}
\begin{figure}
\begin{center}
\includegraphics[scale=.8]{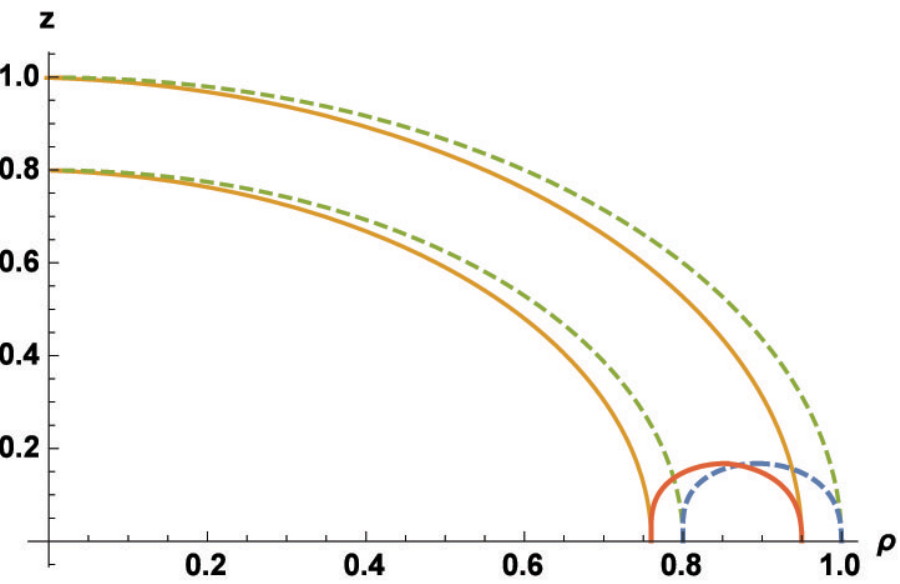}
\end{center}
\caption{Perturbed minimal surfaces versus unperturbed surfaces for $R_-=0.8$, $R_+=1$, $\phi=\pi/2$. The blue dashed curve represents the connected minimal surface and the dashed green ones represent the disconnected minimal surface for $\delta=0$. The red and orange curves represent the connected and disconnected minimal surfaces for $\delta=0.05$.}
\label{fig:AnnProf}
\end{figure}

The solution for the first equation is known to be $\rho_0(z)=\pm\sqrt{R^2-z^2}$ for two symmetric branches. Note that we have ignored $\rho_{22}(z)$ function since it is shown in \cite{Allais:2014ata} that it does not contribute to the first non-trivial correction in entanglement entropy.

\begin{figure}
\begin{center}
\includegraphics[scale=.8]{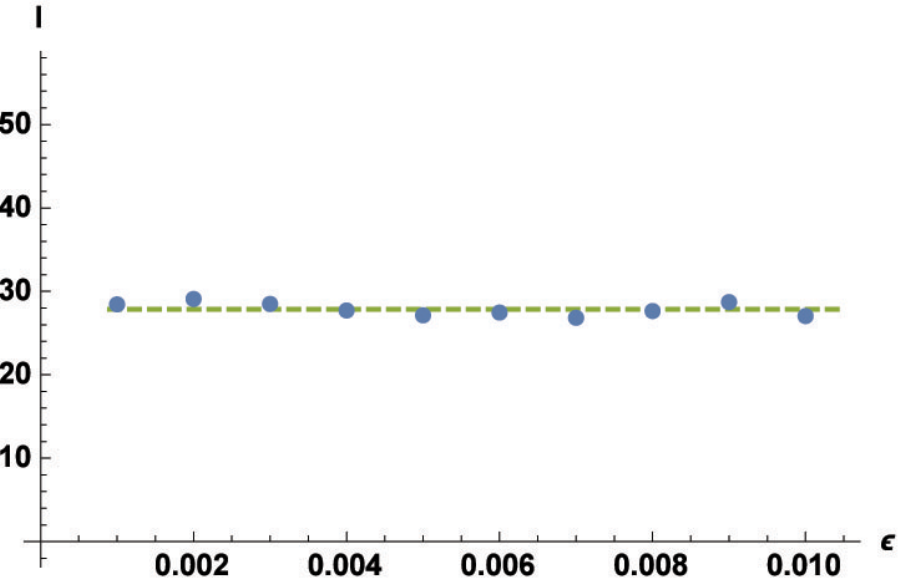}
\hspace{5mm}
\includegraphics[scale=.7]{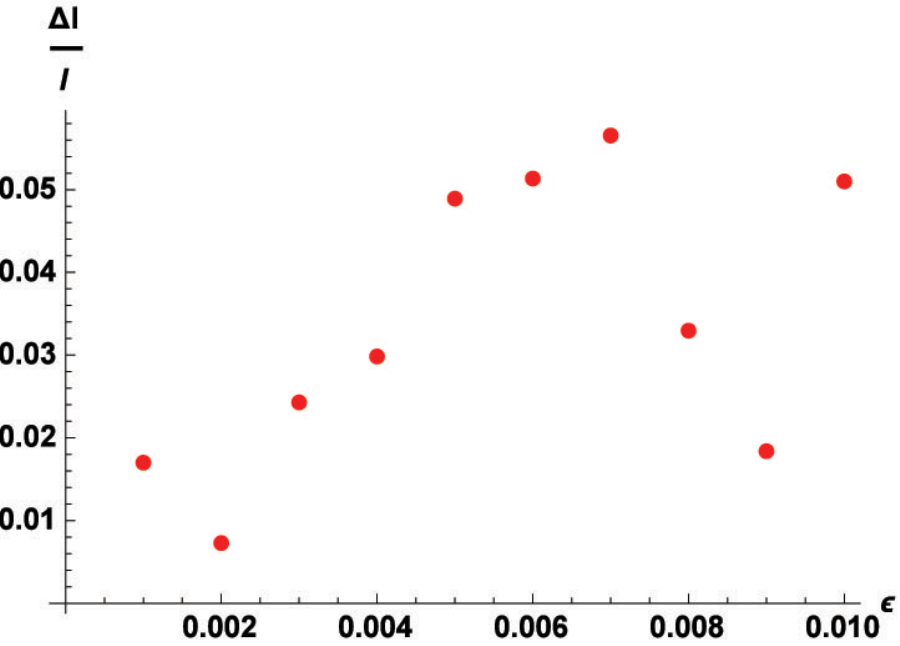}
\end{center}
\caption{Left: The mutual information for the same parameters as a function of $\epsilon$. The dashed green line represents the mean value for mutual information. Right: Contribution of the sharpness to mutual information which is normalized by the mean value of mutual information as a function of $\epsilon$. This shows that the sharpness contribution is independent of the UV cut-off for about $95\%$. We have set $R_-=0.8$, $R_+=1$ and $\delta=0.05$.} 
\label{fig:ShAnn}
\end{figure}

To calculate the mutual information between $A$ and $B$ we have to solve the above equations to find $S_{A}$, $S_{B}$ and $S_{A\cup B}$. To compute $S_{A\cup B}$ we must consider both connected and disconnected configurations which may contribute. By disconnected configurations we mean two minimal surfaces which end on the deformed disks $\rho_{\pm}(\phi)$ which is already considered while we have found $S_{A}$ and $S_{B}$. On the other hand the connected configuration is the minimal surface starting from $\rho_{+}(\phi)$ and ending on $\rho_{-}(\phi)$. Note that the unperturbed connected configuration is also found analytically in \cite{Fonda:2014cca}. We have solved Eq.\eqref{eq:deq} numerically up to $\mathcal{O}\left(\delta^2\right)$ for functions $\rho_0(z)$, $\rho_1(z)$ and $\rho_{20}(z)$ which contribute to the correction of entanglement entropy at first non-trivial order of the sharpness.

In our numerical analysis we have considered $R_-=0.8$, $R_+=1$, $\delta=0.05$ and we have briefly reported the numerical results in Fig.\ref{fig:AnnProf} and Fig.\ref{fig:ShAnn}. In Fig.\ref{fig:AnnProf} we have plotted minimal surface profiles of such a configuration in a constant-$\phi$ slice. The profiles are found by the perturbation theory which is organized such that the turning point of the minimal surfaces are always unchanged ($z_*=.8$ for $R_-$ and $z_*=1$ for $R_+$). In this figure we have compared the profiles for $\delta=0$ and $\delta=0.05$ at $\phi=\pi/2$ slice which has the maximum separation between the perturbed and unperturbed profiles. Note that from the right panel of Fig.\ref{fig:annular} it is obvious that the maximum separation between the deformed and undeformed configurations happens at $\phi=\pi/2$ and $\phi=3\pi/2$ and they coincide at $\phi=0,\pi$. We have computed the mutual information for such configurations and studied its dependence on the inverse UV cut-off $\epsilon$. As we expected our numerical results show that mutual information is independent of the cut-off for about $95\%$. In the left panel of Fig.\ref{fig:ShAnn} we have plotted the mutual information for various values of inverse UV cut-off $\epsilon$. In the left panel we have plotted $\Delta I\equiv I_\delta-I_0$ normalized by the numerical value of $I_\delta$ which shows deviation between $0-5\%$.

In summary in this section we have considered two singular regions which their singularities are apart from each other. We have numerically analysed a specific configuration which the opening angles are equal to each other (see the right panel of Fig.\ref{fig:annular}). It is shown that the mutual information is independent of the UV cut-off to a great precision. As we have explained in the introduction section, because of the extensivity of the coefficient of the logarithmic divergent term, we expect such an independency  even for configurations which the opening angle of the singularities are not the same but the singularities are still apart from each other.

\section{Conclusions and Discussions}

In this paper we have mainly studied corner contributions to certain holographic entanglement measures e.g. mutual and tripartite information for three dimensional CFTs. We have considered two different kinds of singular geometries i.e. a set of disjoint sectors of a single infinite circle (cake slices) which have a contact point and also two sharp concentric circles which are completely disjoint. In particular one of our goals was to explore the role of a singular contact point on the behavior of the corresponding holographic mutual information.

In the case of cake slice entangling regions using the previous results for holographic entanglement entropy for a kink we studied holographic mutual information. We have shown that the corresponding HMI is divergent which is due to the common local region shared among these regions near the contact point. Although the resultant HMI was divergent, our analysis revealed that it exhibits a first order phase transition. In this set-up we also studied the holographic tripartite information which is divergent in contrast with the well-known feature of tripartite information, which is finite even when the subsystems share boundaries. Generalizing these results to higher dimensional holographic CFTs we have explored similar behaviors for mutual and tripartite informations.

In the case of two sharp and completely disjoint concentric circles, by performing a numerical study we have found a finite holographic mutual information. Our numerical results show that mutual information is independent of the UV cut-off with a great accuracy. Comparing this result with the HMI for cake slice entangling regions we conclude that a divergent HMI is a reminiscent of the contact point (infinite local correlations) and it does not have anything to do with the geometric singularity.  

In the following of this section we discuss about some aspects of other entanglement measures associated to a singular surface. We will focus in the first set-up, i.e., cake slice entangling regions which may help us to gain more insights into certain singular surfaces having a contact point.

\subsection*{A Universal Measure}
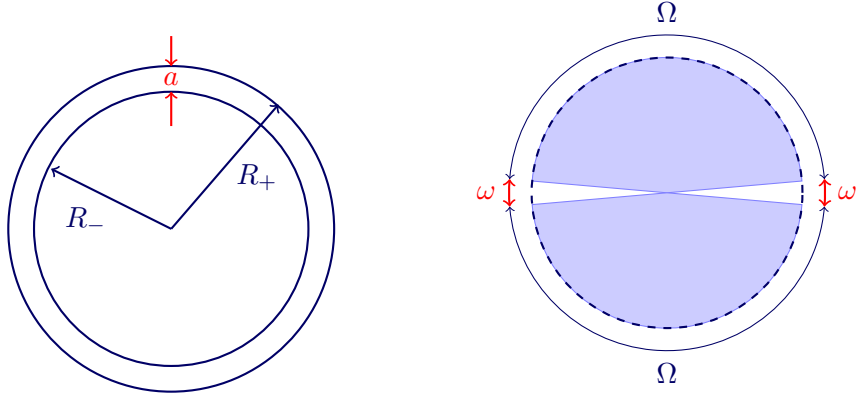
\begin{figure}
\begin{center}
\begin{tikzpicture}[scale=5.7]
\draw[thick,blue!40!black] (-1.5cm,0cm) circle(3.2mm);    
\draw[thick,blue!40!black] (-1.5cm,0cm) circle(3.8mm);    
\draw[blue!40!black,->,thick] (-15mm,0)--(-17.8mm,1.4mm);
\draw[blue!40!black,->,thick] (-15mm,0)--(-12.5mm,2.9mm);
\draw[red,->,thick] (-15mm,2.4mm)--(-15mm,3.2mm);
\draw[red,->,thick] (-15mm,4.5mm)--(-15mm,3.8mm);
%\draw [blue!40!black] (-1.5cm,-1mm) node {$A$};
%\draw [blue!40!black] (-20mm,0mm) node {$B$};
\draw [blue!40!black] (-17mm,.2mm) node {$R_-$};
\draw [blue!40!black] (-13mm,1.2mm) node {$R_+$};
\draw [red] (-1.5cm,3.5mm) node {$a$};
\end{tikzpicture} 
\hspace{1.5cm}
%%%%%%%%%%%%%%%%%%%%%%%%%%%
\begin{tikzpicture}[scale=6]
  \fill[fill=blue!20!white,rotate=5, draw=blue!50!white]
    (0,0) -- (3mm,0mm) arc (0:170:3mm) -- (0,0);
  \fill[fill=blue!20!white,rotate=-5, draw=blue!50!white]
    (0,0) -- (3mm,0mm) arc (0:-170:3mm) -- (0,0);
       \draw[thick,dashed,blue!40!black] (0cm,0cm) circle(3mm);    
%%%%%%%%%%%%%%%%%%
  \draw[blue!40!black,<->, rotate=5]
    (3.5mm,0mm) arc (0:170:3.5mm);
  \draw[blue!40!black,<->, rotate=-5]
    (3.5mm,0mm) arc (0:-170:3.5mm);
  \draw[red!60!red,<->, rotate=-5,thick]
    (3.5mm,0mm) arc (0:10:3.5mm);
  \draw[red!60!red,<->, rotate=175,thick]
    (3.5mm,0mm) arc (0:10:3.5mm);
\draw [blue!40!black] (0,4mm) node {$\Omega$};
\draw [blue!40!black] (0,-4mm) node {$\Omega$};
\draw [red] (4mm,0) node {$\omega$};
\draw [red] (-4mm,0) node {$\omega$};
\end{tikzpicture} 
\end{center}
\caption{Left: The configuration for the mutual information between the disk of Radius $R_-$ and the complement of a disk with radius $R_+$ which is used in \cite{Casini:2015woa} to define a c-function in three dimensions in the $a\to0$ limit. Right: A similar configuration to concentric disks for singular surfaces in the $\omega\to0$ limit for contacting kinks.}
\label{fig:cf}
\end{figure}

In reference \cite{Casini:2015woa}, the authors have introduced a c-function in three dimensional CFTs by means of the mutual information between concentric disks as
\begin{align}
C(R)=\lim_{a\rightarrow 0}\frac{1}{4\pi}\left(R \frac{\partial I}{\partial R}-I\right),
\end{align}
where $a$ is the difference between the radii of the disks (see left panel of Fig.\ref{fig:cf}). In the $a\rightarrow 0$ limit these two disks coincide and they have a common boundary (which leads to infinite MI). As we have shown in section \ref{sec:HEM}, the HMI between sectors of an infinite circle with a contact point at the center of the circle is UV cut-off dependent and thus it is not universal. An important question about such configurations is whether there is any universal measure which we can use for example instead of mutual information and obtain the same physical information?

One may consider a similar set-up to what is done for co-centric disks in the vanishing limit of their separation for our configuration. See the right panel of Fig.\ref{fig:cf} for a visualization. It is important to note that in the limit where the entangling regions are slightly bended, $\Omega \rightarrow \pi$ and $\omega \rightarrow 0$, our entangling regions share boundary which has a similar structure as in \cite{Casini:2015woa}. We define a new quantity from mutual information as
\begin{align}
\mathcal{I}=\lim_{\omega\rightarrow 0}\frac{\omega}{2}\left(\omega \frac{\partial I}{\partial \omega}+I\right),
\end{align}
which captures a universal phase transition. Using the expressions for sharp Eq.\eqref{smallregion} and smooth Eq.\eqref{largeregion} limits one can easily show that at the leading order $\mathcal{I}\approx\kappa$. It would be interesting to further investigate such a quantity and even try to check whether it satisfies the requirements of a c-function.
%corresponding c-function which at the fixed point is equal to $\kappa$.

\subsection*{First Law of HEE and Modular Hamiltonian}
According to the first law of entanglement entropy the variation of the entanglement entropy at leading order is equal to the variation of the modular Hamiltonian, i.e., $\Delta S=\Delta \langle H_{\rm mod.}\rangle$ \cite{Blanco:2013joa}\footnote{See also \cite{{Bhattacharya:2012mi}, {Allahbakhshi:2013rda}}.}. The modular Hamiltonian is a non-local quantity which is defined in terms of the reduced density matrix as $H_{\rm mod.}\sim -\log \rho_{\rm red.}$. In general the computation of this quantity is not an easy task due to its non-locality. However, there are special cases where the modular Hamiltonian becomes local e.g. for spherical entangling regions \cite{Casini:2011kv}. In this case one can find an exact expression for $H$ in terms of an integral over $T_{00}$ (the time-time component of the stress tensor). This shows that using the first law for EE one may extract some perturbative information about the structure of modular Hamiltonian for more general entangling regions. For example the authors of \cite{Blanco:2013joa} have used the first law to show that in addition to $T_{00}$ other spatial components of stress tensor e.g. $T_{ii}$ also contribute to the modular Hamiltonian for strip entangling regions. 

\begin{figure}
\begin{center}
\includegraphics[scale=.6]{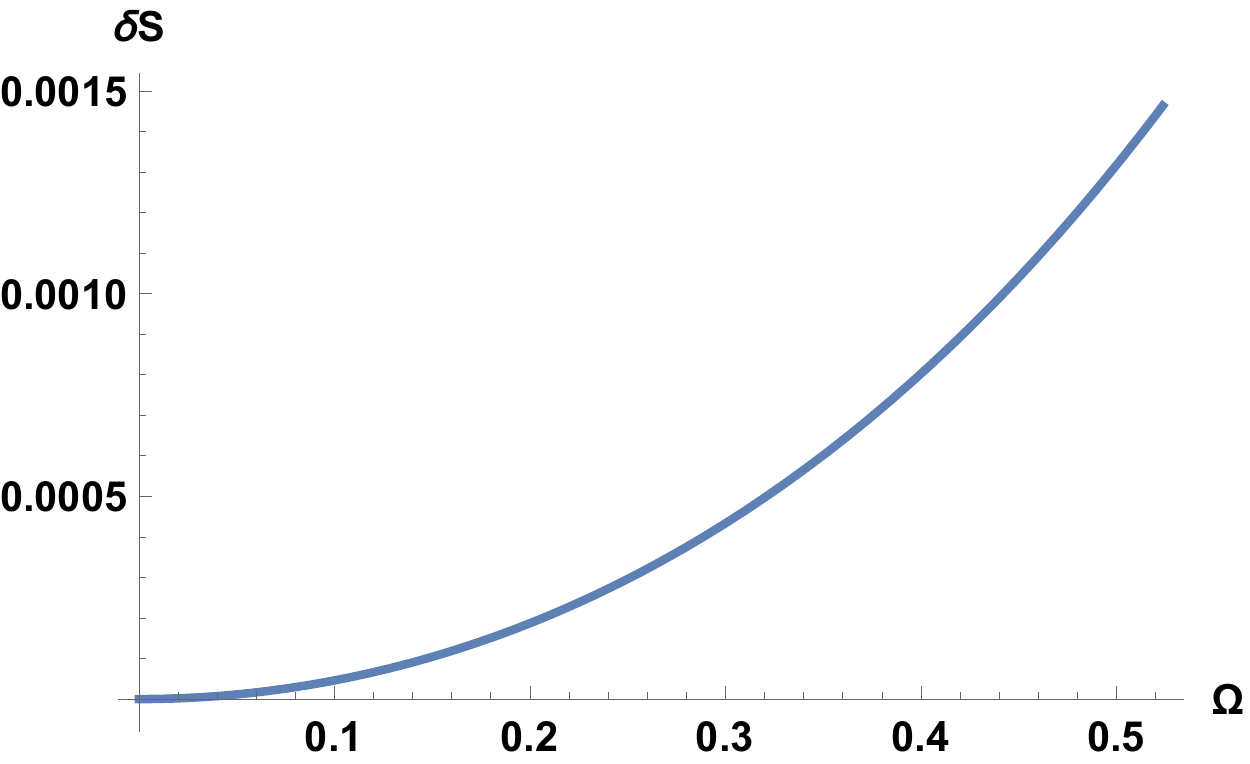}
\end{center}
\caption{Variation of HEE between the ground state (pure AdS) and thermal state (AdS black-brane) in the small opening angle limit for $H=1$ and $m=0.1$.}
\label{fig:deltas}
\end{figure}

In order to find the leading variation of the HEE for a kink entangling region in a three-dimensional holographic CFT we consider an excited state dual to a black-brane back-ground with the following metric
\begin{align}\label{BHmetric}
ds^2=\frac{L^2}{z^2}\left(\frac{dz^2}{f(z)}-f(z)dt^2+d\rho^2+\rho^2 d\theta^2\right),\;\;\;f(z)=1-m z^3.
\end{align}
In this case one may solve the resultant equations for the minimal surface for small entangling regions, i.e., $m h_*^3\ll1$. In Fig.\ref{fig:deltas} we have presented the $\Omega$-dependence of $\Delta S$ in small opening angle limit. 
We expect that this result may help to more investigate corner contributions to modular Hamiltonian which is a nonlocal quantity. 

\section*{Acknowledgements}

The authors would like to thank Pablo Bueno, Piermarco Fonda, Matthew Headrick and specially Horacio Casini for correspondence and valuable comments on an early draft of the manuscript. We are grateful to Mohsen Alishahiha for his valuable comments and careful reading of the manuscript. We also thank Erik Tonni for his correspondence. We would like to thank Pablo Bueno and William Witczak-Krempa for sharing unpublished results on corner and cone entanglement beyond the smooth limit.
FO is very grateful to school of Physics of IPM for its warm hospitality during this project. 
FO would also like to thank the organizers of the PhD school ``Holography: Entangled, Applied, and Generalized" at Niels Bohr Institute in Copenhagen where the final stages of this work took place. 
MRMM and AM are supported by Iran National Science Foundation (INSF). 
\appendix
\section{Holographic Tripartite Information Between Strips}\label{sec:app1}
In this section we aim to show that the holographic prescription of entanglement entropy leads to finite tripartite information even for the case when the disjoint regions merge together and share boundaries. Here we consider an entangling region composed of three disjoint infinite strips (three disjoint segments in a 2-dim field theory) with lengths $\ell_i$ where $i=1,2,3$. These are separated with $h_1$ and $h_2$ as in Fig.\ref{fig:II3}. The tripartite information in this case for a two dimensional field theory by definition is given by
\begin{align}
\begin{split}
I^{(3)}=\log\frac{\ell_1\ell_2\ell_3}{\epsilon^3}
&-\min\left\{\log\frac{\ell_1\ell_2}{\epsilon^2},\log\frac{(\ell_1+\ell_2+h_1)h_1}{\epsilon^2}\right\}\\
&-\min\left\{\log\frac{\ell_2\ell_3}{\epsilon^2},\log\frac{(\ell_2+\ell_3+h_2)h_2}{\epsilon^2}\right\}\\
&-\min\left\{\log\frac{\ell_1\ell_3}{\epsilon^2},\log\frac{(\ell_1+\ell_2+\ell_3+h_1+h_2)(\ell_2+h_1+h_2)}{\epsilon^2}\right\}\\
&+\min\Bigg\{\log\frac{\ell_1\ell_2\ell_3}{\epsilon^3},\log\frac{\ell_3(\ell_1+\ell_2+h_1)h_1}{\epsilon^3},\log\frac{h_1 h_2(\ell_1+\ell_2+\ell_3+h_1+h_2)}{\epsilon^3},\\
&\hspace{10mm}\log\frac{\ell_1(\ell_2+\ell_3+h_2)h_2}{\epsilon^3},\log\frac{\ell_2(\ell_1+\ell_2+\ell_3+h_1+h_2)(\ell_2+h_1+h_2)}{\epsilon^3}\Bigg\}.
\end{split}
\end{align}
There are several cases where this configuration of entanglig regions may share boundaries. By shared boundary we mean that the separation between disjoint subregions `vanishes'. Here `vanishes' precisely means it tends to the value of $\epsilon$. If we consider the entangling regions on a periodic spatial direction, then we denote the spacing between the first and third segments by $h_3$. Now there may be several configurations among $h_i$'s where one, two or even all of them `vanish'. In the following we consider two of these configurations. In the first one $h_1$ `vanishes' while  $h_2$ is finite. The case where $h_2$ `vanishes' and $h_1$ is finite could be obtained by a permutation in the indices. In the second case we consider both $h_1$ and $h_2$ `vanish'.

\begin{figure}
\begin{center}
\begin{tikzpicture}
\draw [blue,thick] (0,0)--(1.3,0);
\draw [densely dashed] (1.3,0)--(2,0);
\draw [blue,thick] (2,0)--(3.8,0);
\draw [densely dashed] (3.8,0)--(4.5,0);
\draw [blue,thick] (4.5,0)--(6,0);
%%%%%%%%%%%%
\draw (.7,-.4) node {$\ell_1$};
\draw (1.7,-.4) node {$h_1$};
\draw (2.9,-.4) node {$\ell_2$};
\draw (4.2,-.4) node {$h_2$};
\draw (5.25,-.4) node {$\ell_3$};
\end{tikzpicture}
\caption{Entangling region with three disjoint parts.}
\label{fig:II3}
\end{center}
\end{figure}

\vspace{3mm}
\textbf{Case 1:}
Consider $\ell_1$ and  $\ell_2$ regions to share a boundary, in other words $h_1\to\epsilon$. In this case together with assuming $h_2,\,\ell_i\gg\epsilon$, one can easily check that the above expression for two dimensional field theory gives
\begin{align}
\begin{split}
I^{(3)}=\log\frac{\ell_2(\ell_1+\ell_2+\ell_3+h_2)}{(\ell_1+\ell_2)(\ell_2+\ell_3+h_2)},
\end{split}
\end{align}
which is independent of the inverse UV cut-off $\epsilon$ as expected. Note that this is in contrast with what happens to mutual information of two disjoint regions in the limit where their distance `vanishes' which is UV cut-off dependent as
\begin{align}
\begin{split}
I=\log\frac{\ell_1\ell_2}{(\ell_1+\ell_2)}+\log\frac{1}{\epsilon}.
\end{split}
\end{align}
A direct generalization of this case to $d>2$ is straightforward where we have to consider infinite strips instead of a line segment. For $d>2$ holographic field theories one can again work out the value of $I^{(3)}$ in the above configuration analytically. For simplicity we consider $\ell_1=\ell_2=\ell_3\equiv\ell$ which leads to
\be
I^{(3)}=-\frac{R^{d-1}}{4G_N^{d+1}}\cdot\frac{2^{d-1}}{d-2}\cdot\frac{\pi^{\frac{d-1}{2}}}{\ell^{d-2}}\cdot\left(\frac{\Gamma\left(\frac{d}{2(d-1)}\right)}{\Gamma\left(\frac{1}{2(d-1)}\right)}\right)^{d-1}\cdot f_d
\ee
where
\be
f_d=
\begin{cases}
1+3^{2-d}-2^{3-d}&~~ h_2^{d-2}\left(2^{d-1}-1\right)<\left(2\ell\right)^{d-2}\\
\left(\frac{\ell}{h_2}\right)^{d-2}-1+3^{2-d}-2^{2-d} &~~ h_2^{d-2}\left(2^{d-1}-1\right)\ge \left(2\ell\right)^{d-2}
\end{cases}.
\ee
Again note that the mutual information of two infinite strips when their separation `vanishes' is again a UV cut-off dependent quantity as
\be
I=\frac{R^{d-1}}{4G_N^{d+1}}\cdot\frac{2^{d-1}}{d-2}\cdot\frac{\pi^{\frac{d-1}{2}}}{\epsilon^{d-2}}\cdot\left(\frac{\Gamma\left(\frac{d}{2(d-1)}\right)}{\Gamma\left(\frac{1}{2(d-1)}\right)}\right)^{d-1}+\mathcal{O}\left(\epsilon^0\right).
\ee

\textbf{Case 2:}
One can easily find the result where both $h_1$ and $h_2$ `vanish' simultaneously from the results of case 1 for holographic field theories. For the case of $d=2$ one can easily check that the tripartite information is again a UV cut-off independent quantity as
\begin{align}
\begin{split}
I^{(3)}=\log\frac{\ell_2(\ell_1+\ell_2+\ell_3)}{(\ell_1+\ell_2)(\ell_2+\ell_3)},
\end{split}
\end{align}
and for the case of $d>2$ it becomes
\be
I^{(3)}=-\frac{R^{d-1}}{4G_N^{d+1}}\cdot\frac{2^{d-1}\left(1+3^{2-d}-2^{3-d}\right)}{d-2}\cdot\frac{\pi^{\frac{d-1}{2}}}{\ell^{d-2}}\cdot\left(\frac{\Gamma\left(\frac{d}{2(d-1)}\right)}{\Gamma\left(\frac{1}{2(d-1)}\right)}\right)^{d-1}
\ee

\section{Hyperscaling-violating Geometries}\label{sec:Hs}
\begin{figure}
\begin{center}
\includegraphics[scale=.7]{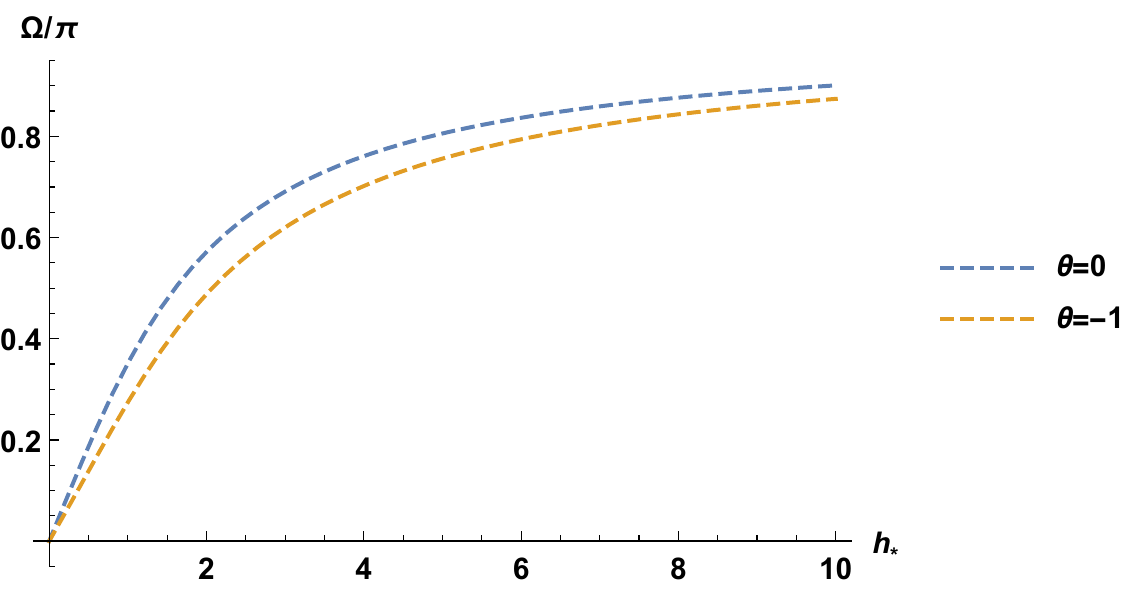}
\includegraphics[scale=.7]{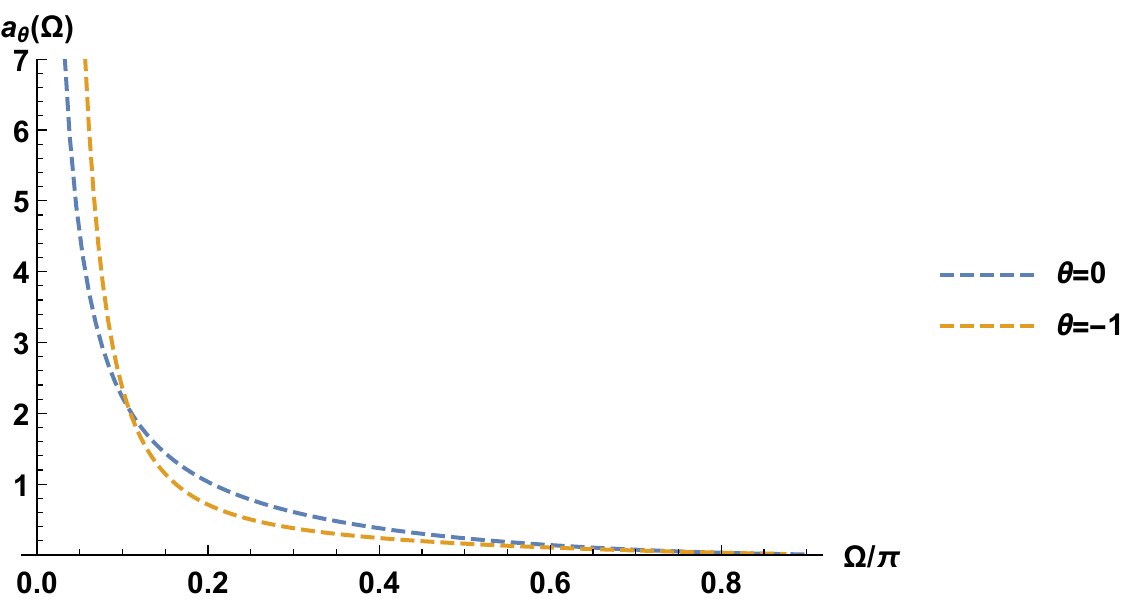}
\end{center}
\caption{\textit{Left}: $\Omega/\pi$ as a function of the turning point $h_*$. \textit{Right}: $a_{\theta}$ as a function of the opening angle $\Omega$.}
\label{fig:aomega-hyper}
\end{figure}

In this appendix we generalise our studies to theories with a hyperscaling-violating geometry as a gravity dual. The HEE for a singular entangling region in this geometry has been previously studied in \cite{Alishahiha:2015goa}. Here we explore holographic mutual information in specific examples of this kind of backgrounds. As in section \ref{sec:HEM} we only consider crease entangling regions and set the dynamical critical exponent $z=1$. The metric for a hyperscaling-violating geometry in four dimensions is given by \cite{Charmousis:2010zz}
\begin{align}\label{metrichyper}
ds^2=\frac{1}{z^{2-\theta}}\left(dz^2-dt^2+d\rho^2+\rho^2 d\phi^2\right),
\end{align}
where $\theta$ is the hyperscaling violating exponent and due to the null energy condition it must satisfy either $\theta\geq 2$ or $\theta\leq 0$ \cite{Dong:2012se}. Also in order to avoid gravitational instabilities we only consider $\theta\leq 0$ in the following discussion \cite{Dong:2012se}. Considering the entangling region as in Eq.\eqref{entangling} and rewriting 
$z(\rho,\phi)=\rho\;h(\phi)$ such that $h(\pm\frac{\Omega}{2})=0$, the HEE functional becomes 
\begin{align}\label{heefunc-hyper}
S=\frac{L^2}{2G_N}\int_{\frac{\epsilon}{h_*}}^H \frac{d\rho}{\rho^{1-\theta}}\int_0^{\frac{\Omega}{2}-\delta}d\phi\frac{\sqrt{1+h^2+h'^2}}{h^{2-\theta}},
\end{align}
where the notations are similar to the previous sections. Since the integrand does not depend on $\phi$ explicitly we can define a conserved quantity such that
\begin{align}\label{thetamom-hyper}
\mathcal{H}_\theta\equiv \frac{(1+h^2)^{\frac{2-\theta}{2}}}{h^{2-\theta}\sqrt{1+h^2+h'^2}}=\frac{(1+h_*^2)^{\frac{1-\theta}{2}}}{h_*^{2-\theta}}.
\end{align}
Using this expression one can find the relation between the opening angle $\Omega$ and the turning point in the bulk as follows
\begin{align}\label{omehah0-hyper}
\Omega =2\int_0^{h_*} \frac{dh}{\sqrt{1+h^2}\sqrt{(\frac{h_*}{h})^{2(2-\theta)}(\frac{1+h^2}{1+h_*^2})^{1-\theta}-1}}
\end{align}
and if we use the change of variable $y=\sqrt{\frac{1}{h^2}-\frac{1}{h_*^2}}$ the HEE becomes
\begin{align}\label{HEE-hyper}
S(\Omega)=\frac{L^2}{2G_N}\int_{\frac{\epsilon}{h_*}}^H \frac{d\rho}{\rho^{1-\theta}}\int_0^{\sqrt{\frac{\rho^2}{\epsilon^2}-\frac{1}{h_*^2}}}dy \frac{h_* y \left(\frac{1}{h_*^2}+y^2\right)^{-\frac{\theta }{2}}}{\sqrt{1+h_*^2 y^2-\left(\frac{h_*^2+1}{h_*^2 \left(y^2+1\right)+1}\right)^{1-\theta }}}.
\end{align}
The behaviour of the integrand near the boundary, i.e., $y\rightarrow \infty$ is given by
\begin{align}
\frac{h_* y \left(\frac{1}{h_*^2}+y^2\right)^{-\frac{\theta }{2}}}{\sqrt{1+h_*^2 y^2-\left(\frac{h_*^2+1}{h_*^2 \left(y^2+1\right)+1}\right)^{1-\theta }}}\sim 
\begin{cases}
y+\mathcal{O}(\frac{1}{y^5})&\theta=-1,\\
y^{-\theta}+\frac{\#}{y^{\theta+2}}+\mathcal{O}(\frac{1}{y^{\theta+4}})&\theta\leq -2,
\end{cases}
\end{align}
which shows that in order to isolate the divergent part one needs to apply a $\theta$-dependent regularization. For example when $\theta =-1$ one finds
\begin{align}\label{HEEhyperfinal}
S(\Omega)=\frac{L^2}{4G_N}\left(\frac{H}{\epsilon^2}-\frac{2}{h_*\epsilon}+\frac{1}{h_*^2H}\right)+a_{-1}(\Omega)\left(\frac{h_*}{\epsilon}-\frac{1}{H}\right)+\mathcal{O}(\epsilon),
\end{align}
where the function $a_{-1}(\Omega)$ is defined as
\begin{align}
a_{-1}(\Omega)=\frac{L^2}{2G_N}\int_0^\infty dy\;y\left[\frac{ \left(1+h_*^2y^2\right)^{\frac{1}{2}}}{\sqrt{1+h_*^2 y^2-\left(\frac{h_*^2+1}{h_*^2 \left(y^2+1\right)+1}\right)^{2}}}-1\right].
\end{align}
A similar procedure leads to the divergent parts for other values of $\theta$. Fig.\ref{fig:aomega-hyper} demonstrates the behavior of $\Omega(h_*)$ and $a_{\theta}(\Omega)$ for $\theta=-1$. In this figure we have also included the case of $\theta=0$ for comparison. Also note that the function $a_{\theta}(\Omega)$ vanishes in $\Omega\rightarrow \pi$ limit which corresponds to the smooth limit as expected. According to Eq.\eqref{HEEhyperfinal} even in the smooth limit in addition to the area law term there exists an extra divergent term. Explicit computations show that when one computes the HMI this divergent term does not cancel out and the resultant HMI becomes divergent. This situation is similar to the case of a spherical entangling region in a $d$-dimensional field theory with $d>3$.
 
\end{document}